\documentclass{emulateapj}
\usepackage{graphicx,amsmath,amsfonts,amssymb,subfigure}

\slugcomment{Accepted by ApJ}

\begin{document}

\title{Tracing galaxies through cosmic time with number density selection}

\author{Joel Leja\altaffilmark{1},  Pieter van Dokkum\altaffilmark{1}, Marijn Franx\altaffilmark{2}}

\altaffiltext{1}{Astronomy
  Department, Yale University, New Haven, CT~06520.}
\altaffiltext{2}{Sterrewacht Leiden, Leiden University, NL-2300 RA Leiden,
Netherlands}

\begin{abstract}
\renewcommand{\thefootnote}{\fnsymbol{footnote}}
A central challenge in observational studies of galaxy formation is how to associate progenitor galaxies with their descendants at lower redshifts. One promising approach is to link galaxies at fixed number density, rather than fixed luminosity or mass. This method is effective if stellar mass rank order is broadly conserved through cosmic time. In this paper, we use the {Guo} {et~al.} (2011) semi-analytical model to analyze under what circumstances this assumption is valid in the context of a cosmological simulation. Specifically, we select progenitor galaxies at a constant number density and compare the stellar mass evolution of their descendants to the evolution at a constant number density. The median stellar mass of the descendants increases by a factor of four (0.6 dex) from $z = 3$ to $z = 0$. Constant number density selection reproduces this to within 40\% (0.15 dex) over a wide range of number densities. We show that the discrepancy primarily results from scatter in the stellar mass growth rates and merging. After applying simple, observationally-based corrections for these processes, the discrepancy is reduced to 12\% (0.05 dex). We conclude that number density selection can be used to predict the median descendant mass of high-redshift progenitor galaxies. The main uncertainty in this study is that semi-analytical models do not reproduce the observed mass evolution of galaxies, which makes the quantitative aggregate effects of star formation, merging, and quenching on the rank order of galaxies somewhat uncertain.
\end{abstract}


%%%%%%%%%%%%%%%%%%%%%%%%%%%%
%%%%%%%%%%% INTRODUCTION %%%%%%%%%%
%%%%%%%%%%%%%%%%%%%%%%%%%%%%

\section{Introduction}
In order to infer changes in a galaxy population, it is necessary to link progenitor and descendant galaxies at different snapshots in redshift. Yet in order to properly link these galaxies, it is also necessary to know how their observational properties change through time-- i.e., it is necessary to {\it already} have a consistent theory of galaxy evolution. This creates a pseudo "chicken and egg" problem which has been a long-standing challenge for studies of galaxy evolution.

One common solution to the progenitor problem is to link galaxies at fixed stellar mass at different redshifts. This is effective when studying galaxies which are believed to evolve passively, as their properties are relatively constant with time: for example, luminous red galaxies (LRGs) ({Wake} {et~al.} 2006). This method, however, suffers from {\it progenitor bias} ({van Dokkum} \& {Franx} 1996): by only studying galaxies at high redshift that appear similar to local LRGs, some of the progenitors of local LRGs will be missed. Linking galaxies in this fashion will thus systematically underestimate their true evolution. As the lookback time increases, the error introduced by comparing galaxies at fixed mass will also increase, making the method unsuitable for studies beyond the nearby Universe.

Some recent studies ({Wuyts} {et~al.} 2011; {Leitner} 2012) suggest linking local starforming galaxies to their progenitors at high redshift by following the mass evolution implied by the starforming sequence. The tight relationship between stellar mass and star formation rate makes this a powerful technique. However, for this technique to work, the mass added via mergers must be negligible. Additionally, systematic uncertainties in the star formation rate (SFR) are considerable and vary substantially between SFR indicators ({Wuyts} {et~al.} 2011). The technique also cannot be applied to quiescent galaxies, as their quenching redshift and subsequent merger history remains unknown. Since many galaxies are quiescent ({Brammer} {et~al.} 2011), other methods to link descendants and progenitors must be developed.

Tracking galaxies at constant number density is a promising technique that may prove more effective than previous methods. The basic approach, introduced in {van Dokkum} {et~al.} (2010), is to assume that the number density of galaxies when ordered by some physical property (e.g., stellar mass) does not evolve. This provides a simple way to link descendant and progenitor populations at different redshifts. This technique has already been used to study the evolution of a number of different galaxy properties: UV luminosity and star formation history ({Papovich} {et~al.} 2011), the stellar velocity dispersion function ({Bezanson} {et~al.} 2011), H$\alpha$ equivalent width ({Fumagalli} {et~al.} 2012), and mass and structural evolution ({Bezanson} {et~al.} 2009; {van Dokkum} {et~al.} 2010; {Brammer} {et~al.} 2011; {Patel} {et~al.} 2012).

However, investigations into the effectiveness of the technique have been limited. The investigations that do exist rely primarily on numerical results. Van Dokkum (2010) used a simple Monte Carlo simulation in their appendix to show that mergers have little effect on the stellar mass evolution measured at a constant number density. {Papovich} {et~al.} (2011) went slightly further, using halo merger trees from the Millenium Simulation ({Springel} {et~al.} 2005) to show the recovery fraction of true descendant halos from z$\sim$7 to z$\sim$3 is approximately 50\%. Dark matter halo mass evolves in a fundamentally different way than stellar mass, though, and can be subject to substantial tidal stripping when a halo becomes a satellite: this makes it less suitable for galaxy tracking than stellar mass. No study to date has attempted to examine the efficacy of number density selection in a more realistic simulation setting. Additionally, the theoretical underpinnings of number density selection remain unexplored: the rules governing how it works, why it works, and where it breaks down remain an open question.

This paper will study number density selection by first discussing the assumptions inherent in the use of number density selection and which physical processes might violate them. The {Guo} {et~al.} (2011) semi-analytical model (SAM) applied to the Millenium Simulation is then used to investigate the validity of these assumptions. Semi-analytical models represent our best theoretical formulation of how galaxy evolution works on a broad scale, and simulations are currently the only way in which it is possible to track the actual evolution of individual galaxies. In addition, it is not necessary for our purposes that the SAM accurately reproduces the observed Universe. This makes the SAM useful as a simple tool to explore the mechanics of number density selection and what processes may cause it to err.

In Section 2, we introduce the number density selection technique and discuss potential sources of error. In Section 3, we describe the Millenium Simulation, the {Guo} {et~al.} (2011) semi-analytical model, and describe how number density selection is applied to these data. In Section 4, we present the resulting median stellar mass evolution and several techniques to improve it. In Section 5, we explore the possibility of ordering galaxies by inferred velocity dispersion rather than stellar mass. In Section 6, the suitability of applying SAM-based results to the real Universe is discussed. The conclusions are found in Section 7.

When necessary, we assume H$_0$=70 km s$^{-1}$ Mpc$^{-3}$ throughout the paper.

%%%%%%%%%%%%%%%%%%%%%%%%%%%%%%%%%%%%%%%%%%%%%%%%
%%%%%%%%%%% CONSTANT NUMBER DENSITY SELECTION %%%%%%%%%%%%%
%%%%%%%%%%%%%%%%%%%%%%%%%%%%%%%%%%%%%%%%%%%%%%%%

%%%%%%%
%% FIG 1 %%
%%%%%%%

\begin{figure}[t!]
\begin{center}
\includegraphics[scale=0.45]{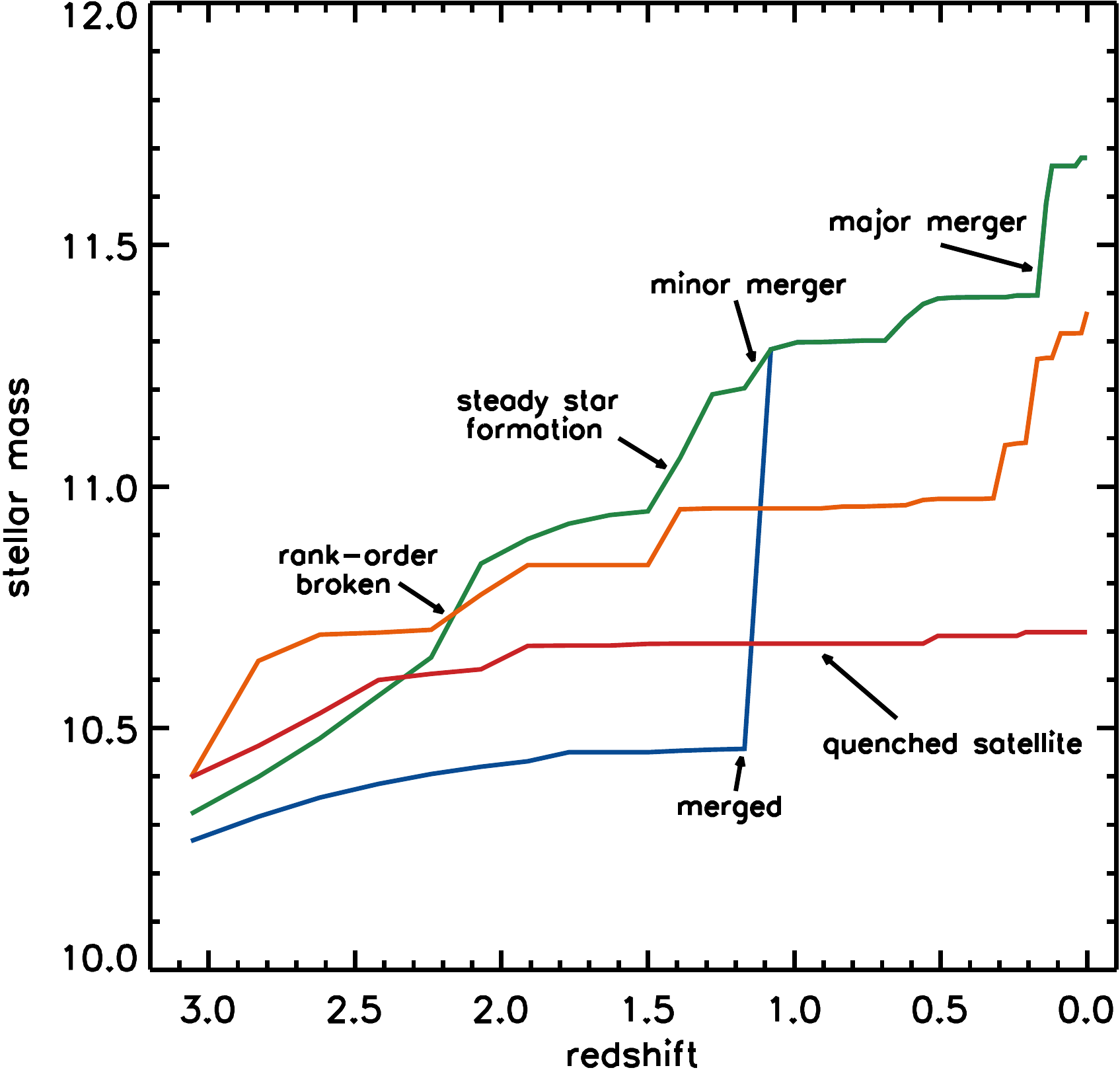}
\caption{Stellar mass assembly histories are shown for four representative galaxies from the Guo et al. (2011) semi-analytical model output. Despite similar initial stellar masses, their mass growth histories vary substantially. The primary mechanisms for stellar mass growth (major and minor mergers, star formation) are labelled. Furthermore, the processes interfering with number density selection-- scatter in growth rate, merged galaxies, and quenching-- are also labelled. In order to for constant number density selection to produce robust results, these processes must be either negligible or correctable.}
\label{galtraj}
\end{center}
\end{figure}

\section{Constant Number Density Selection}
We discuss the underlying assumptions and potential pitfalls of the number density selection technique here. As described earlier, linking progenitors and descendants at a constant number density implicitly assumes that the number density of galaxies does not evolve. One can then infer the mass evolution of a given galaxy population from the redshift evolution of the galaxy stellar mass function. Number density selection thus provides a physically-motivated, redshift dependent stellar mass selection. The preferred approach is to link galaxies at a constant {\it cumulative} number density, which has the advantage of being single-valued in mass ({Brammer} {et~al.} 2011). Additionally, the physical interpretation is simple: if the stellar mass rank order of galaxies remains the same through cosmic time, then tracking galaxies at a cumulative number density will be effective.

The aim of this study, then, is to investigate the conditions under which the conservation of rank order holds. If stellar mass growth is either constant or monotonically increasing in proportion to stellar mass itself, then rank order will be preserved. The stellar mass\footnote{Note that stellar mass in the Guo et al (2011) model refers to the integral of the star formation rate. {Guo} {et~al.} (2011) do not explicitly include mass loss from SNe and winds in their model; instead a mass fraction of 0.43 is immediately returned to the ISM.} of a galaxy can be increased in two ways: star formation and merging. Steady star formation in starforming galaxies is proportional to M$_{*}^{\alpha}$ (where $\alpha > 0$) over a wide range of redshifts, commonly referred to as the starforming sequence ({Brinchmann} {et~al.} 2004; {Noeske} {et~al.} 2007; {Peng} {et~al.} 2010; {Gonzalez} {et~al.} 2012; {Whitaker} {et~al.} 2012). The galaxy-galaxy merger rate is also expected, on average, to increase monotonically with stellar mass (see, e.g., the review by {Hopkins} {et~al.} (2010)). Thus, there is good reason to believe rank order in stellar mass may be conserved.

However, there also exist physical processes which will confuse rank order. Measured scatter in the starforming sequence is 0.34 dex; this includes intrinsic scatter from AGN, starbursts, and galaxies in the process of shutting down their star formation as well as measurement error ({Whitaker} {et~al.} 2012). The stochastic nature of mergers naturally induces scatter in stellar mass growth over short timescales. These two effects in combination are referred to as the scatter in the stellar mass growth rate, and can break rank order by either "jumping" a galaxy ahead or "leaving" a galaxy behind. An example of breaking rank order by scatter in growth rates is shown in Figure \ref{galtraj} by the green galaxy trajectory.

Galaxy mergers will also change the cumulative number density by decreasing the total number of galaxies above a fixed stellar mass. This will cause a constant number density line to underestimate the true median mass evolution. This issue is only relevant for mergers with galaxies that are at least as massive as the galaxy under consideration. As lower-mass galaxies have a higher number density than the number density under consideration, they do not affect the inferred stellar mass evolution.

For completeness, one must also consider the creation of new galaxies. The stellar mass at which this effect is important is well below current detection limits in high-redshift surveys. For example, the mass-complete limit for the Newfirm Medium Band Survey ({Whitaker} {et~al.} 2011) is M$_*$ $>$ 10$^{11}$ M$_{\odot}$ (all galaxies) or M$_*$ $\gtrsim 10^{10}$ M$_{\odot}$ (starforming galaxies) at z=2.2 ({Brammer} {et~al.} 2011). Assuming newly formed galaxies at this epoch lie on the starforming sequence, they would not approach the mass limit in any reasonable amount of time. The creation of new galaxies may then be safely neglected.

Finally, quenching may significantly affect rank order. Quenched galaxies no longer grow via star formation, whereas their star-forming companions at a similar mass will continue to add stellar mass. For central galaxies, this may be offset somewhat as they still experience significant growth through mergers: see Section \ref{quenching} for further discussion. Redshift evolution of the quenching rate and quenching mass limit may also confuse the rank order of galaxies.

The effects of these processes on the rank order of galaxies are illustrated in Figure \ref{galtraj}, and must be well characterized in order to safely use number density selection to track galaxy evolution.

%%%%%%%%%%%%%%%%%%%%%%%%%%%%%%%%%%%%%%%%%%%%%%%%
%%%%%%%%%%% SIMULATION DATA AND SELECTION %%%%%%%%%%%%%
%%%%%%%%%%%%%%%%%%%%%%%%%%%%%%%%%%%%%%%%%%%%%%%%

%%%%%%%
%% FIG 2 %%
%%%%%%%

\begin{figure}[t!]
\begin{center}
\includegraphics[scale=0.4]{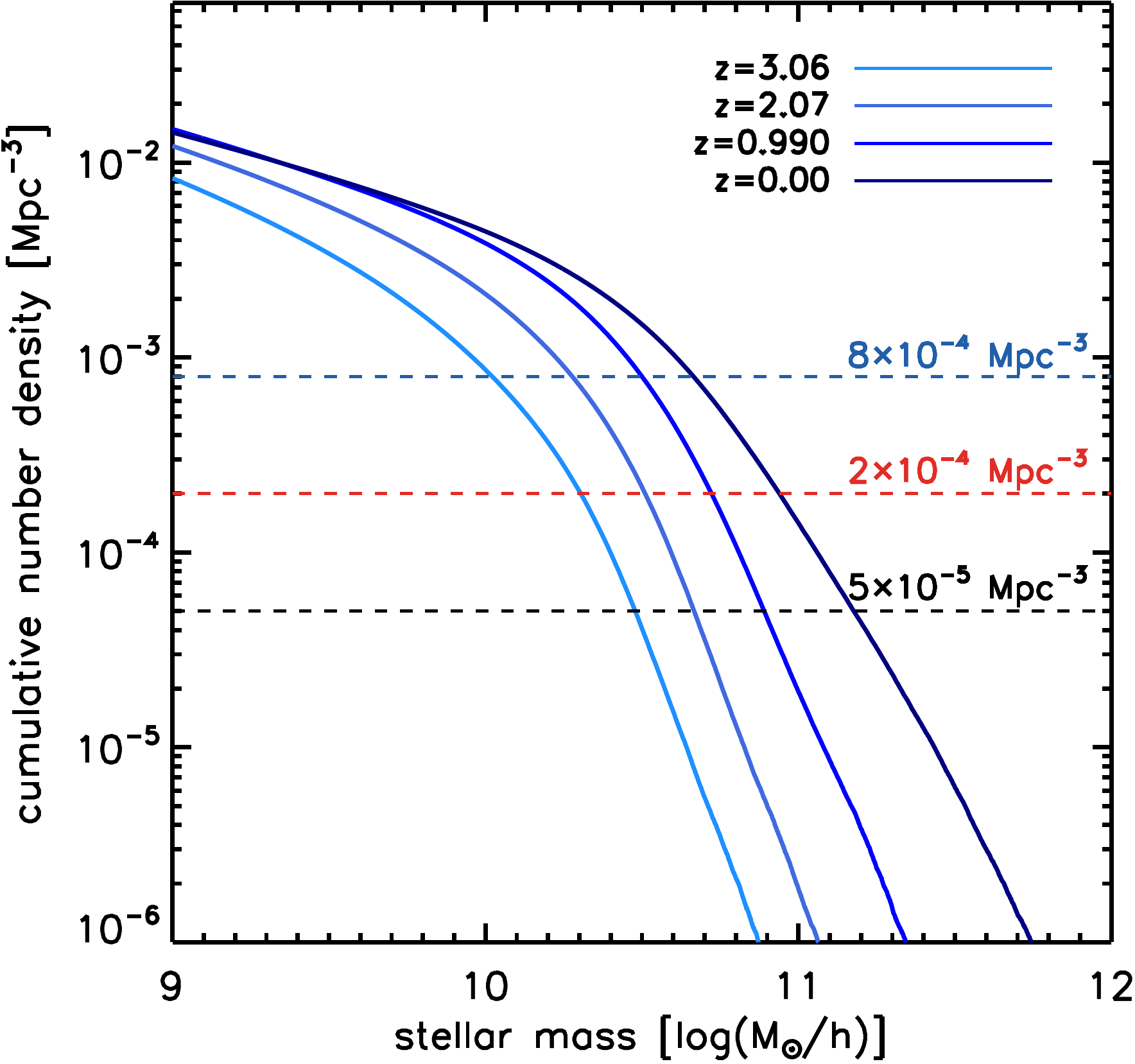}
\includegraphics[scale=0.4]{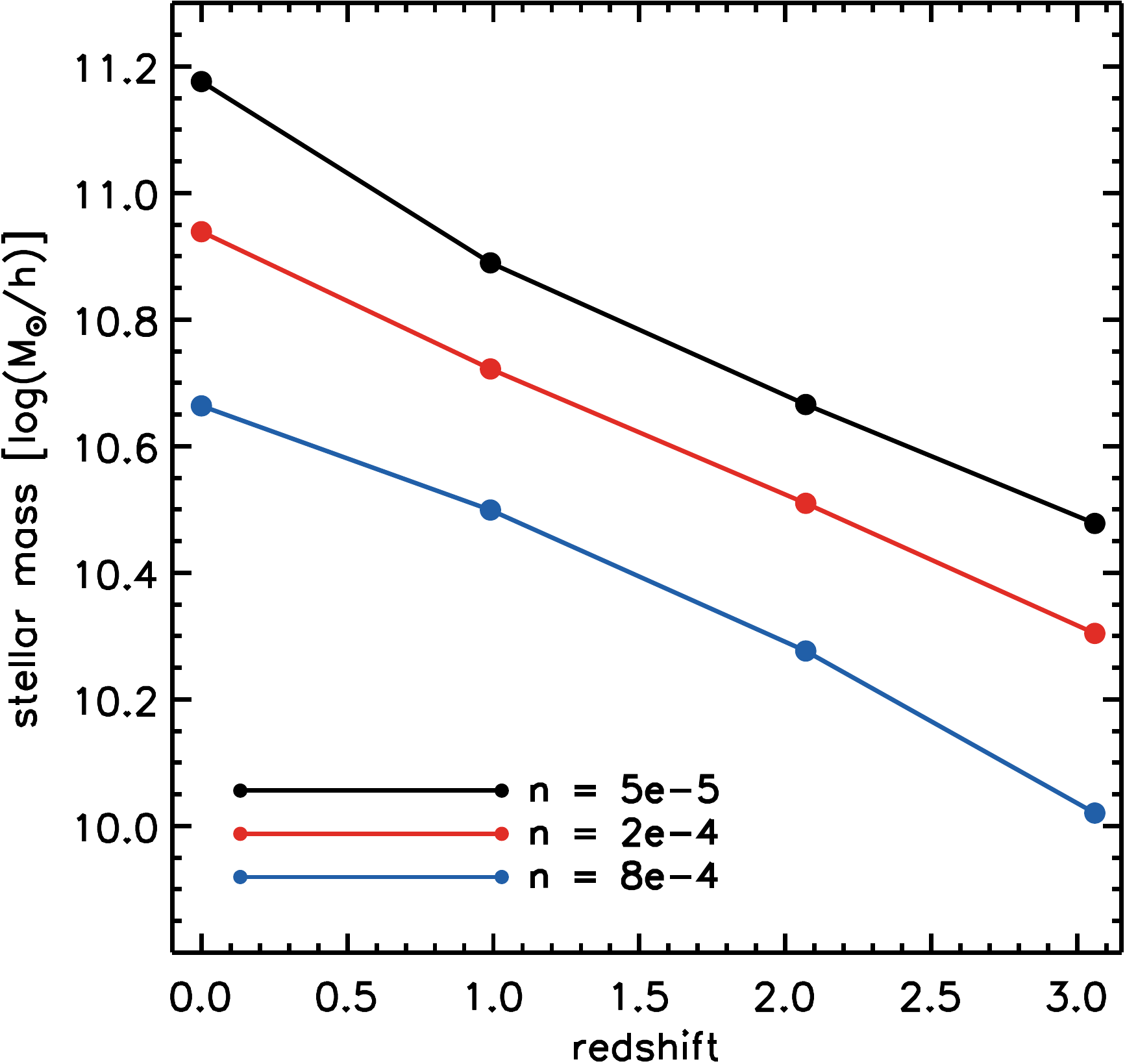}
\caption{The stellar mass function for four redshift slices from the {Guo} {et~al.} (2011) mock galaxy catalogs is illustrated here. The three number density slices investigated in this paper are shown as dotted lines. The intersection of the constant number density lines with the mass functions represent the expected mass evolution of the number density selected sample. This expected median mass evolution is also shown here as a function of number density.}
\label{numdens}
\end{center}
\end{figure}

\section{Simulation Data and Selection}
\subsection{The Millenium Simulation}\label{sec:mill}
We test the number density selection technique using mock galaxy catalogs from the {Guo} {et~al.} (2011) semi-analytical model, which is based on N-body output from the Millenium Simulation ({Springel} {et~al.} 2005). We access these data using the online relational database built by the German Astrophysical Virtual Observatory ({Lemson} \& {Virgo Consortium} 2006).

The Millenium Simulation is a large cosmological N-body simulation following the evolution of 2160$^3$ particles from $z=127$ to the present. The Millenium cosmology is based on a joint analysis of the 2dFGRS ({Colless} {et~al.} 2001) and WMAP1 results ({Spergel} {et~al.} 2003). The input cosmological parameters are $\Omega_{\Lambda}$ = 0.75, $\Omega_m$ = 0.25, $\Omega_b = 0.045$, $n=1$, $\sigma_8 = 0.9$, and $H_0 = 73$ km s$^{-1}$ Mpc$^{-1}$. While most of these parameters have changed in the recent WMAP7 results ({Komatsu} 2010), the deviation is small enough to have little effect on the results ({Guo} {et~al.} 2011). The only significant deviation is in $\sigma_8$, which is almost 4$\sigma$ off the WMAP7 value ($\sigma_8 = 0.809\pm0.024$). The effect of an inflated $\sigma_8$ parameter on merger rates and galaxy catalogs will be discussed in Section \ref{sec:cosmo}. 

The Millenium Simulation is run in a periodic box 500h$^{-1}$ Mpc on a side with a particle mass of $8.6 \times 10^{8}  $h$^{-1}$ M$_{\odot}$. Post-processing is performed to separate particles into friends-of-friends groups ({Davis} {et~al.} 1985). The SUBFIND algorithm ({Springel} {et~al.} 2001) is then used to split friends-of-friends groups into bound subhalos. The largest subhalo is referred to as the main subhalo, and contains most of the mass. A subhalo must contain at least twenty particles to be identified by the SUBFIND algorithm, which corresponds to a minimum subhalo mass of $1.72 \times 10^{10}$ M$_{\odot}$.

\subsection{The Guo et al. (2011) SAM}\label{SAM}
The {Guo} {et~al.} (2011) SAM is an updated version of the earlier Millenium-based galaxy formation models ({De Lucia} {et~al.} 2006). It successfully matches the abundance of $\sim$10$^{11}$ M$_{\odot}$ galaxies out to $z\sim3$. Additionally, it accurately reproduces the local stellar mass function by construction. However, due to the inflated $\sigma_8$ parameter in the Millenium cosmology, galaxies cluster more strongly than in the real Universe. This results in a high fraction of quenched red satellites. In order to match the local mass function, the SAM must assemble these low-mass quenched satellites early in the simulation. Thus, the abundance of low-mass galaxies beyond $z=0.6$ is overpredicted. The effect of this on constant number density selection is discussed in Section \ref{discussion}.

Additionally, due to the finite resolution of the N-body simulation, the mock galaxy catalogs are only reliable above a certain stellar mass limit. Application of the SAM to the higher-resolution Millenium II indicates that the mass functions and morphological classifications begin to diverge substantially around $\log{M_{star}} \sim 9.8$ at $z=0$ ({Guo} {et~al.} 2011). Number density slices are chosen carefully so as to track galaxies only when above this stellar mass limit (see Section \ref{nummill}). Galaxy evolution subsequent to crossing the resolution limit is expected to behave with more regularity.
\subsection{Number Density Selection in the Guo et al. (2011) SAM}\label{nummill}

\begin{table}[t!]
\begin{center}
\caption{Number Density Samples}
\begin{tabular}{cccc}
\tableline
\tableline
\multicolumn{1}{c}{n [Mpc$^{-3}$]}
 & \multicolumn{1}{c}{$\Delta$n}
 & \multicolumn{1}{c}{log(M$_{*}/h$) [z=3]} \\
 \tableline
$5 \times 10^{-5} $ & $5 \times 10^{-5}$ & 10.48\\
$2 \times 10^{-4} $ & $1 \times 10^{-4}$ & 10.31\\
$8 \times 10^{-4} $ & $1 \times 10^{-4}$ & 10.02\\
\tableline
\tableline
\end{tabular}
\end{center}
\end{table}

Here we discuss the implementation of number density selection within the SAM.

First, the mock galaxy catalog is used to build stellar mass functions at the redshift outputs nearest to $z= 3$, $z=2$, $z=1$, and $z=0$ (see Figure \ref{numdens}). This covers the typical range of observational studies and includes the last $\sim$11 Gyr of galaxy formation. Galaxies at three different number densities are selected at $z=3$, and their descendants are followed until $z=0$. The mass evolution of these descendants is compared to the mass evolution inferred at a constant number density. We note that measuring the stellar mass at a fixed number density makes no use of a bin.

Merging not only grows the stellar mass of galaxies, but also decreases the total number of galaxies. In order to self-consistently count galaxies, galaxies that merge with a more massive galaxy are considered to have "disappeared", and are discarded from the sample. As a consequence of this, the number of galaxies descended from the $z=3$ number density bin will decrease as time passes. The magnitude of this effect is discussed in Section \ref{sec:merge}.

The chosen number density slices are shown in Table 1, along with the initial bin width at $z=3$ and the corresponding mass at constant number density. The number density slices were chosen to sample a variety of stellar masses while remaining above the stellar mass resolution limit of $\log{M_{star}} \sim 9.8$ (see the discussion in Section \ref{SAM}). The bin size is constant in number density rather than in stellar mass for two reasons: (1) the median stellar mass is invariant with respect to bin size for a bin in number density, and (2) a bin size in number density provides a self-consistent way to compare bin size if one orders galaxies by a different parameter, as is done in Section \ref{sec:disp}.. Additionally, a bin size specified in units of cumulative number density means the projected number density of selected galaxies in constant in each redshift interval, which may be useful in observational studies.

\begin{figure}[t]
\begin{center}
\includegraphics[scale=0.4]{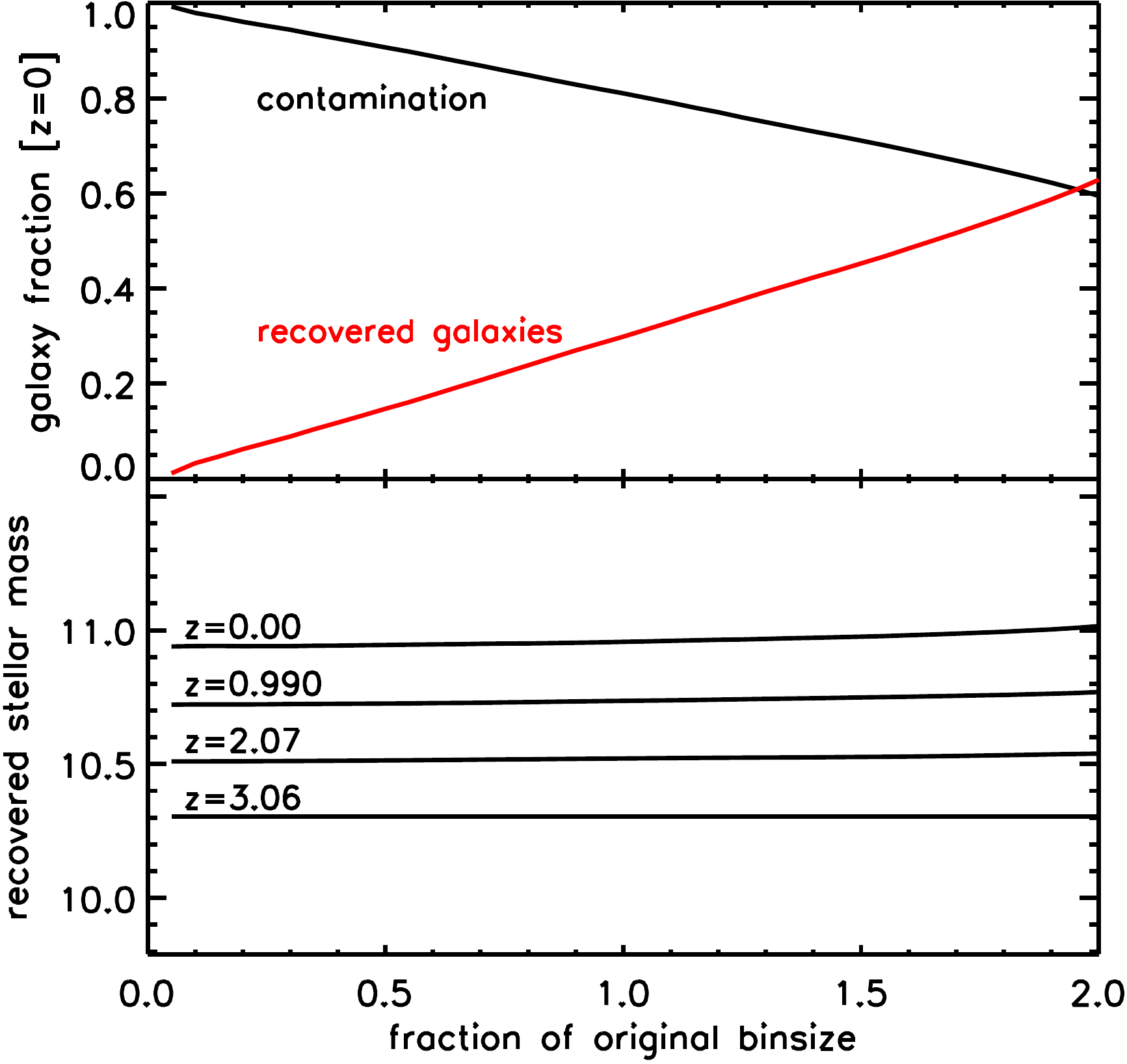}
\caption{Examining the effect of varying bin size with the $n = 2 \times 10^{-4}$ Mpc$^{-3}$ slice in number density. The recovery fraction is defined as the number of descendant galaxies within the bin divided by the total number of surviving descendant galaxies, while the contamination fraction is defined as the number of galaxies in the bin {\it not} descended from the original selection divided by the total number of galaxies within the number density bin. Both are shown in the upper panel. The lower panel shows the change in the measured median stellar mass of the descendants. Bin size has very little effect on the recovered stellar mass, whereas recovery fraction and contamination fraction are strong functions of the chosen bin size. This suggests that recovery fraction and contamination fraction poor diagnostics to test number density selection, whereas median stellar mass is a much more robust indicator.}
\label{binsize}
\end{center}
\end{figure}

In order to ensure the results of this study are independent of the method of investigation, the bin size is varied for the $n=2\times10^{-4}$ Mpc$^{-3}$ slice. The resulting effect on descendant mass evolution is shown in Figure \ref{binsize}. The descendant mass evolution is very stable with respect to the bin size. Also shown is the recovery fraction, defined as the number of descendant galaxies within the bin divided by the total number of surviving descendant galaxies, and the contamination fraction, defined as the number of galaxies in the bin {\it not} descended from the original selection divided by the total number of galaxies within the number density bin. The recovery and contamination fractions are a function of (1) the median mass growth rate, (2) the dispersion in mass growth rate, and (3) the initial bin size. While these diagnostics have figured prominently in previous studies of number density selection [e.g. {van Dokkum} {et~al.} (2010); {Papovich} {et~al.} (2011)], they are not discussed in this paper as they are demonstrated to be a strong function of bin size. Additionally, it is not necessary to recover the exact same galaxies with number density selection as long as the recovered galaxies are indistinguishable from the true descendants in important galaxy properties such as stellar mass.

%%%%%%%
%% FIG 3 %%
%%%%%%%

%%%%%%%%%%%%%%%%%%%%%%%%%%%%%%%%%%%%%%%%%%%%%%%%%%%%
%%%%%%%%%%% TRACKING THE MASS EVOLUTION OF DESCENDANTS %%%%%%%%%%%%
%%%%%%%%%%%%%%%%%%%%%%%%%%%%%%%%%%%%%%%%%%%%%%%%%%%%

\section{Tracking Stellar Mass Evolution}

%%%%%%%
%% FIG 4 %%
%%%%%%%

\begin{figure}[t]
\begin{center}
\includegraphics[scale=0.4]{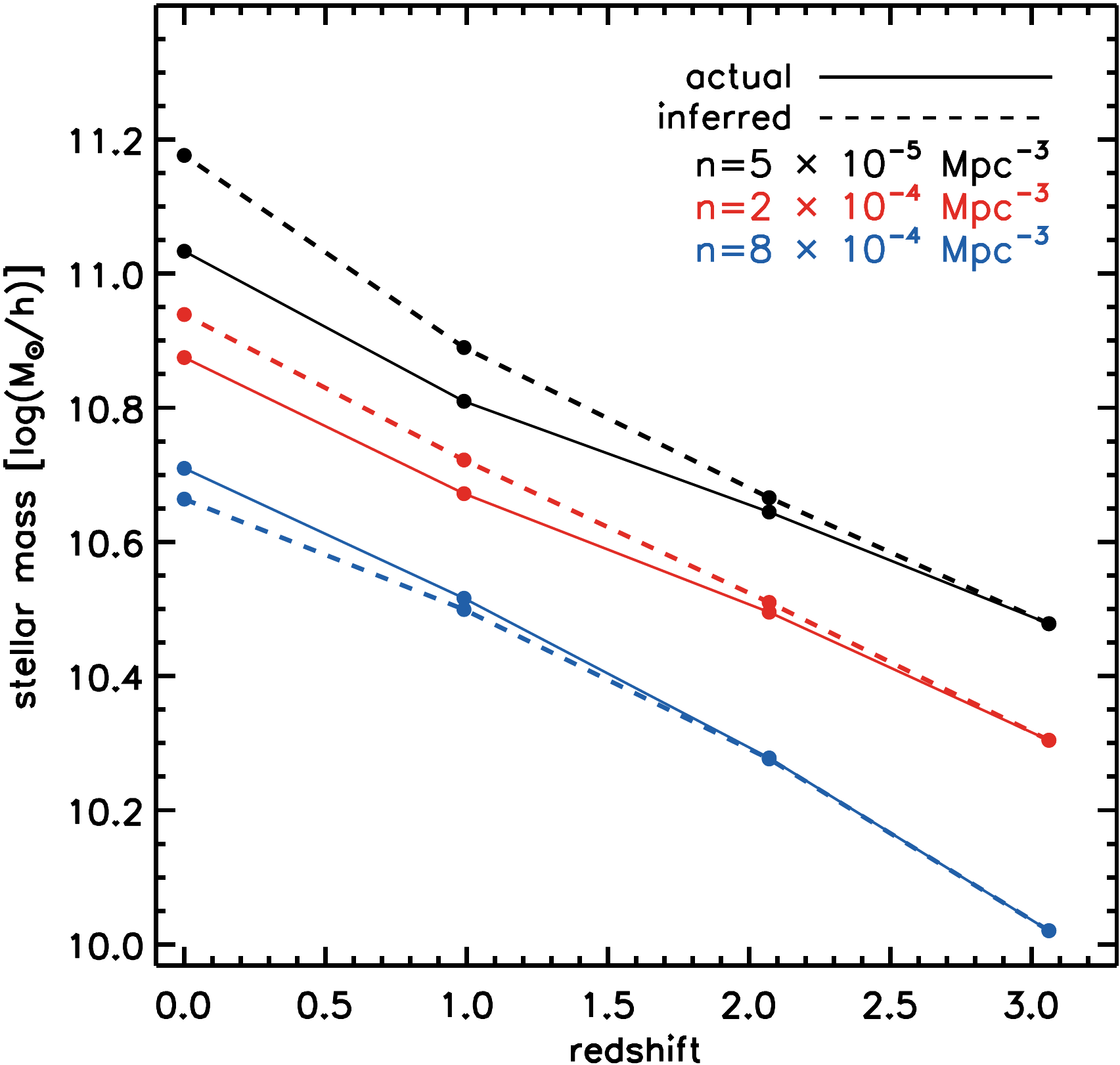}
\caption{The median mass growth of the true descendants (solid line) compared to the mass growth inferred at a constant number density (dashed line) for three bins in number density. The mass evolution is reproduced correctly to within 40\% (0.15 dex) over an evolution of $\sim$4 in mass (0.6 dex) and 11 Gyr in time. Comparing galaxies at the same stellar mass would be off by a factor of four (0.6 dex). Note that the mass evolution of the highest number density sample is systematically underpredicted, while the mass evolution of the two lower number density samples are systematically overpredicted. This can be explained by the scatter in growth rates and the merging of galaxies, as described in Sections 4.1 and 4.2, respectively.}
\label{simpmass}
\end{center}
\end{figure}

We follow the median stellar mass of galaxies from the original $z=3$ sample and compare it to the stellar mass evolution at a constant cumulative number density. Galaxies that merge with a more massive galaxy are removed from the sample, as described in Section \ref{nummill}. The results are shown in Figure \ref{simpmass}.

The median stellar mass of the descendants is reproduced to within 0.15 dex over 0.6 dex of evolution in mass and $\sim 11$ Gyr in time for all number density slices. This corresponds to a factor of four growth in stellar mass predicted to within 40\%. Excluding the $z=0$ point for the lowest number density bin, the error decreases to 20\%, or 0.08 dex. As time passes, the discrepancy between the predicted evolution and the actual evolution increases. This is because processes which scramble rank order have more time to operate.

The direction of the discrepancy between the median descendant stellar mass and the stellar mass at a constant number density is consistent within each number density sample. The two lower number density samples overpredict the mass evolution of the descendants while the highest number density sample underpredicts the mass evolution of the descendants. The physical processes responsible for this are identified in Figures \ref{specevo} and \ref{mergfrac}, and corrected for in Figure \ref{stellgrowth_numdens}. Discussion of these procedues is found in the following sections.
\subsection{Scatter in Growth Rates}
\label{subsec:scattersec}

%%%%%%%
%% FIG 5 %%
%%%%%%%
\begin{figure}[t]
\begin{center}
\includegraphics[scale=0.4]{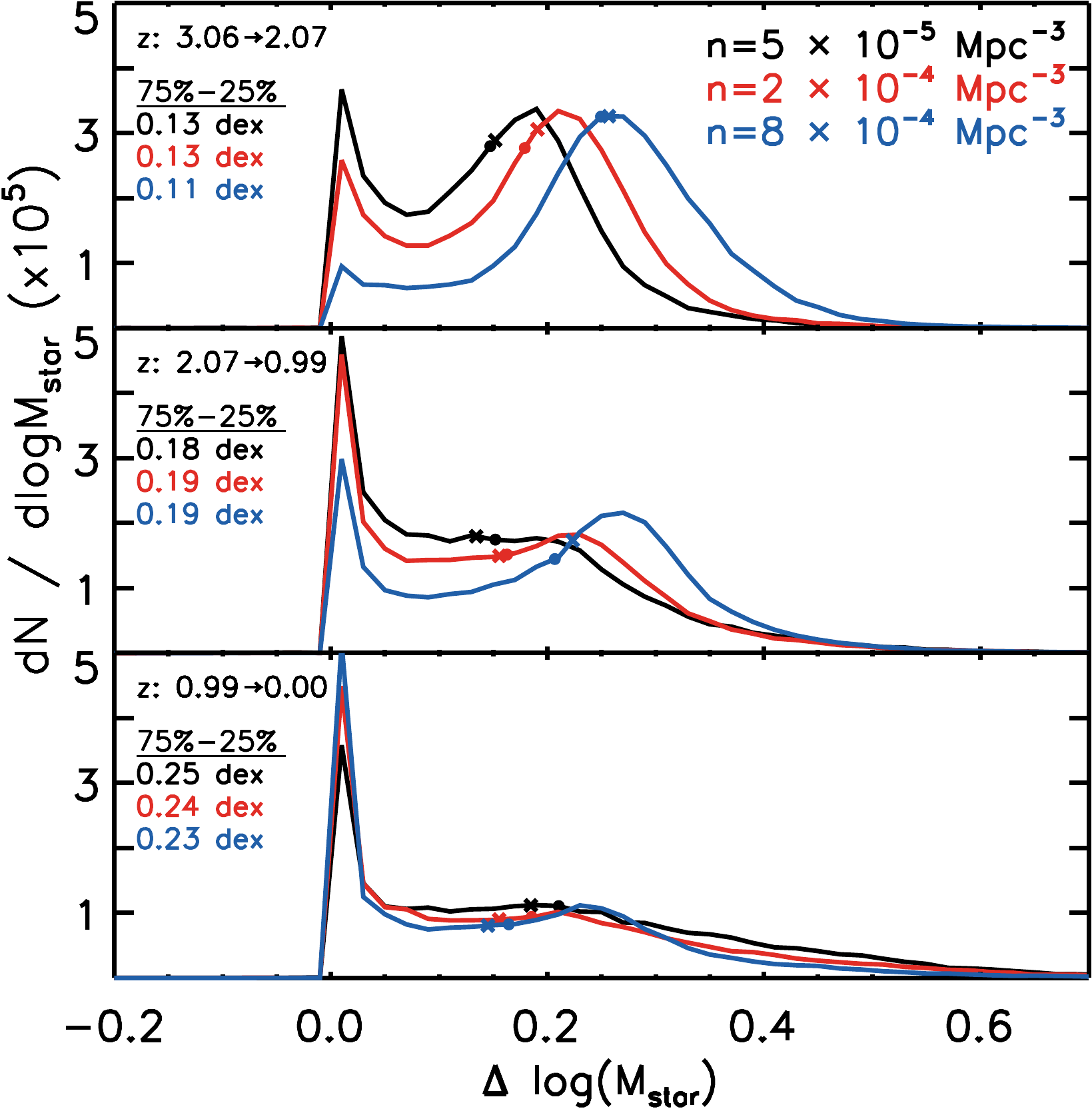}
\caption{Tracing the evolution of descendants from the initial selection at $z=3.06$. The panels show the distribution of evolution in stellar mass for individual galaxies. The average is marked with a circle, while the median is marked with a cross. The difference between the 75th quartile and the 25th quartile is shown in the upper left of each panel. The $n=5\times10^{-5}$ Mpc$^{-3}$ descendants are multiplied by a factor of two for comparison, to compensate for the smaller initial bin size.}
\label{specevo}
\end{center}
\end{figure}

The two lower number density bins consistently overpredict the mass evolution of the sample. One plausible explanation for this is the scatter in stellar mass growth rates. As the stellar mass function is steep for these low number densities, there is a large reservoir of slightly lower mass galaxies. Due to the scatter in growth rates, some of these low mass galaxies will grow faster than the descendants of the $z=3$ sample, overtaking them in stellar mass: this is a version of the well-known Eddington bias ({Eddington} 1913). To investigate this possibility, the distribution of actual growth rates for individual galaxies in each number density bin is shown in Figure \ref{specevo}.

There is a peak at null growth in each number density bin. This peak is populated with galaxies that experienced no star formation nor significant growth via mergers over the time period. This peak is initially largest for the lowest number density population. However, as gas supplies dwindle, more and more galaxies become quenched. By $z=0$, the highest number density bin has the largest number of galaxies that exhibit little to no growth. This is likely because the lower number density bins contain more massive centrals which are still experiencing growth via merging.

There is also a secondary peak in each growth rate which is most evident at the highest redshift. This is the peak due to steady star formation from galaxies on the star-forming sequence. Its width is a combination of the finite size of the mass bin and scatter in the star forming sequence. As more galaxies quench, this peak washes out.

The median mass growth is marked with a cross. Initially, the highest number density galaxies experience the largest relative growth. Note that this is not in violation of the assumption that more massive galaxies grow more quickly: in linear units, the lowest number density bin adds more mass than the higher number density bins. By the final redshift bin, as merging begins to dominate the stellar mass growth, this has reversed, and the highest number density galaxies grow at a faster relative rate than the lower number density galaxies.

An estimate of the effect of scatter in growth rates on number density selection can be made by convolving the stellar mass functions shown in Figure \ref{numdens} with the individual growth rates shown in Figure \ref{specevo}. At each snapshot in redshift, the growth rate for each number density slice is convolved with the stellar mass function at that redshift. This will "grow" each galaxy in mass with a probability distribution given by Figure \ref{specevo}. Due to the shape of the mass function, the convolved stellar mass function results in a larger inferred mass growth than the unconvolved stellar mass function for all three number densities. The difference between the mass growth inferred from the convolved mass function and that inferred from the original mass function is taken to be the error in inferred growth rate for each number density slice. The inferred mass growth adjusted for this effect is shown in the bottom left panel of Figure \ref{stellgrowth_numdens}.

This procedure is only an approximation, as it assumes the measured growth rate within a number density slice is applicable to all galaxies in the SAM. However, this is a relatively reasonable approximation, as the galaxies that do "overtake" more massive galaxies are likely to be fairly close in stellar mass.  As shown in Figure \ref{stellgrowth_numdens}, the correction successfully removes much of the discrepancy between the inferred stellar mass and actual stellar mass at the lower number densities.
\subsection{Merging Galaxies}\label{sec:merge}
%%%%%%%
%% FIG 6 %%
%%%%%%%
\begin{figure}[t]
\begin{center}
\includegraphics[scale=0.4]{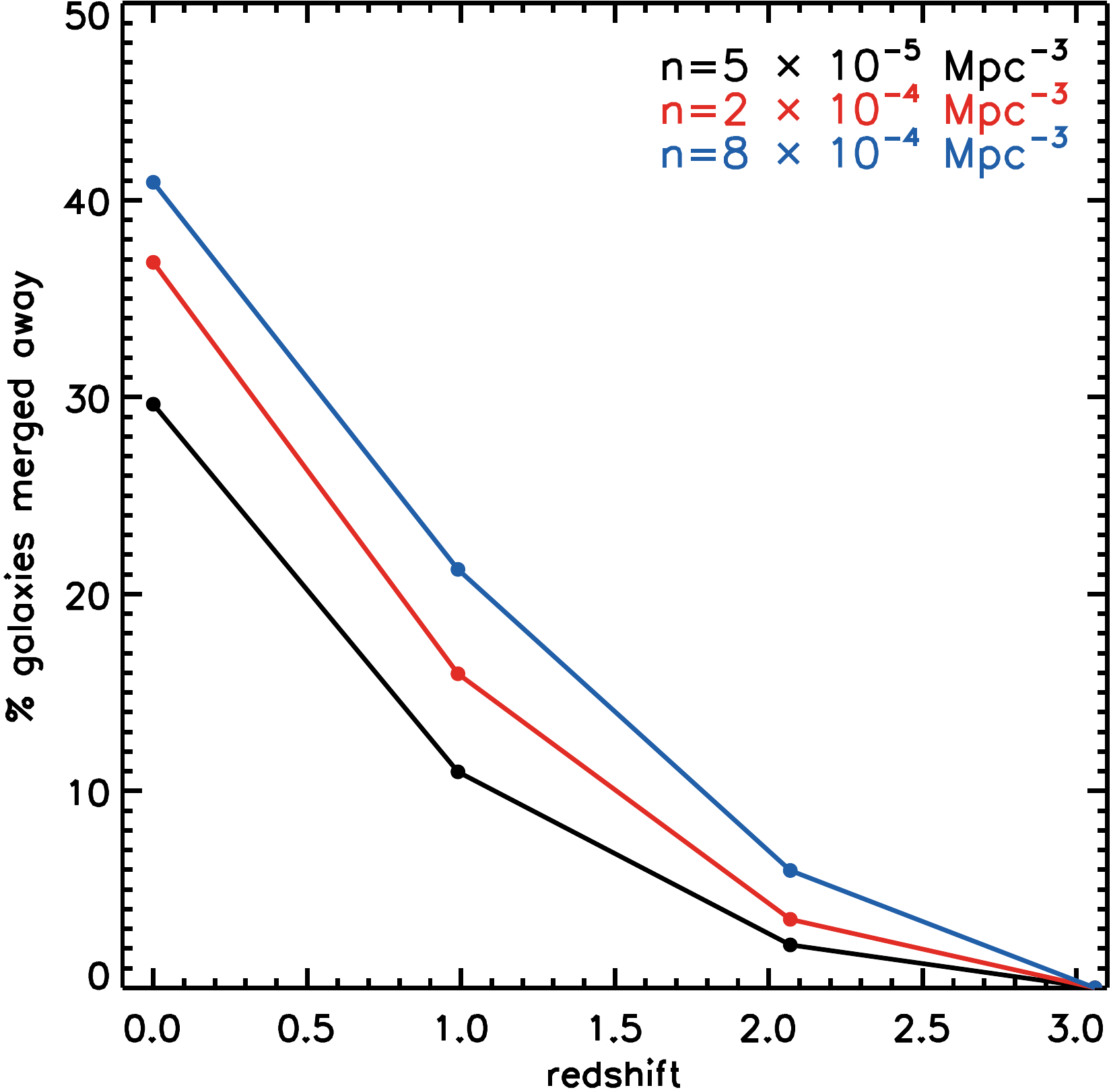}
\caption{The cumulative percentage of galaxies that have lost their identity by merging into a more massive galaxy is shown. The percentage is plotted as a function of number density and redshift. The fraction of merged galaxies increases with number density, as galaxies at a lower number density have few galaxies more massive with which to merge.}
\label{mergfrac}
\end{center}
\end{figure}

The highest number density bin, in contrast, systematically underpredicts the stellar mass evolution of the true descendants. This effect becomes more pronounced after adjusting for the scatter in growth rates. A likely explanation is the change in cumulative number density caused by galaxies that "disappear" due to merger events, as described in Section \ref{nummill}. When a galaxy with stellar mass greater than the sample disappears in a merger, this will decrease the actual cumulative number density of the sample. This results in systematic underprediction of the stellar mass evolution.

\begin{figure*}[t!]
\begin{center}
\includegraphics[scale=0.5]{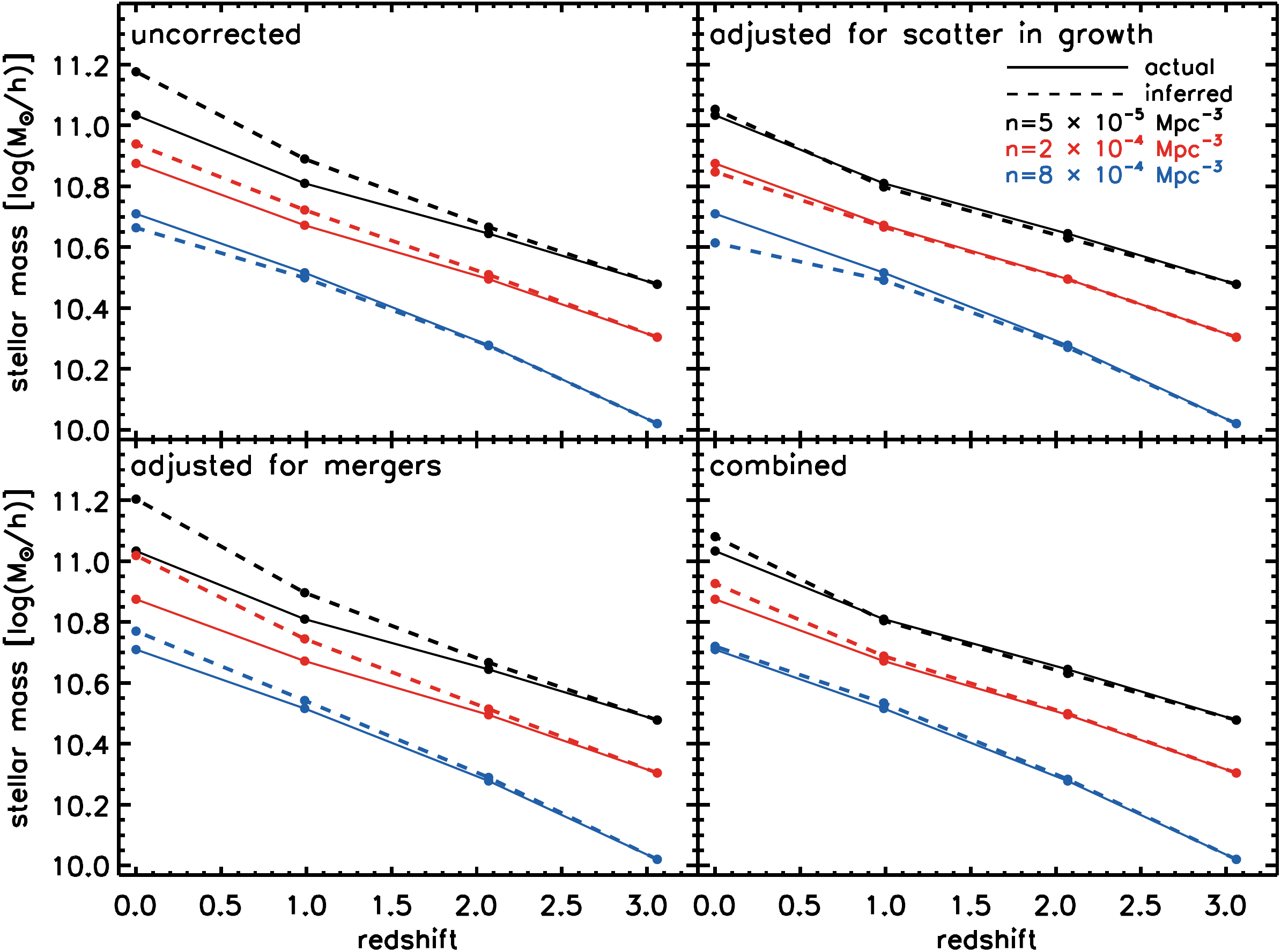}
\caption{The median stellar mass of the true descendants and the corresponding stellar mass inferred from the stellar mass function is shown in three different number density bins. The top left panel shows the median stellar mass with no adjustments. The top right panel shows the median stellar mass after correcting the cumulative number density for galaxies that have been destroyed via mergers. The lower left panel shows the median stellar mass after adjusting the number density selection to accommodate for the scatter in stellar mass growth. The final panel shows the combination of these two corrections. The remaining error in the inferred mass evolution is likely due to a combination of the effects of quenched satellites and the approximate nature of the corrections.}
\label{stellgrowth_numdens}
\end{center}
\end{figure*}

The magnitude of this effect can be calculated exactly if the rate of galaxy-galaxy mergers as a function of stellar mass is known. A simple estimate can be made by calculating the fraction of galaxies that disappear due to mergers as a function of redshift, shown in Figure \ref{mergfrac}. The cumulative number density is multiplied by the merger fraction. As the cumulative number density is decreased, the inferred mass growth will increase. The inferred mass growth corrected for merged galaxies is shown in the top right panel of Figure \ref{stellgrowth_numdens}.

This correction is only approximate, as it implicitly assumes that the merger fraction measured at a given number density holds for all galaxies at lower number densities. A more precise way to correct for this effect would be to integrate the merger fraction rate as a function of number density. This would be a small adjustment for the higher two number density bins, as can be seen in Figure \ref{mergfrac}, since the merger fraction is relatively constant over number density in this regime. The lowest number density bin includes the most massive galaxies in the simulation, however, where the merger fraction must necessarily go to zero at the massive end. It is likely then that this approximation overadjusts the mass evolution in the lowest number density bin. The fact that the fully "corrected" inferred mass growth is larger than the actual mass growth lends support to this hypothesis.

The combined effect of both corrections is shown in the lower right panel of Figure \ref{stellgrowth_numdens}. The agreement is now excellent, with a discrepancy of only 12\% (0.05 dex) over a factor of four (0.6 dex) evolution in stellar mass.

%%%%%%%
%% FIG 7 %%
%%%%%%%

\subsection{Quenching}\label{quenching}
\begin{figure}[t]
\begin{center}
\includegraphics[scale=0.4]{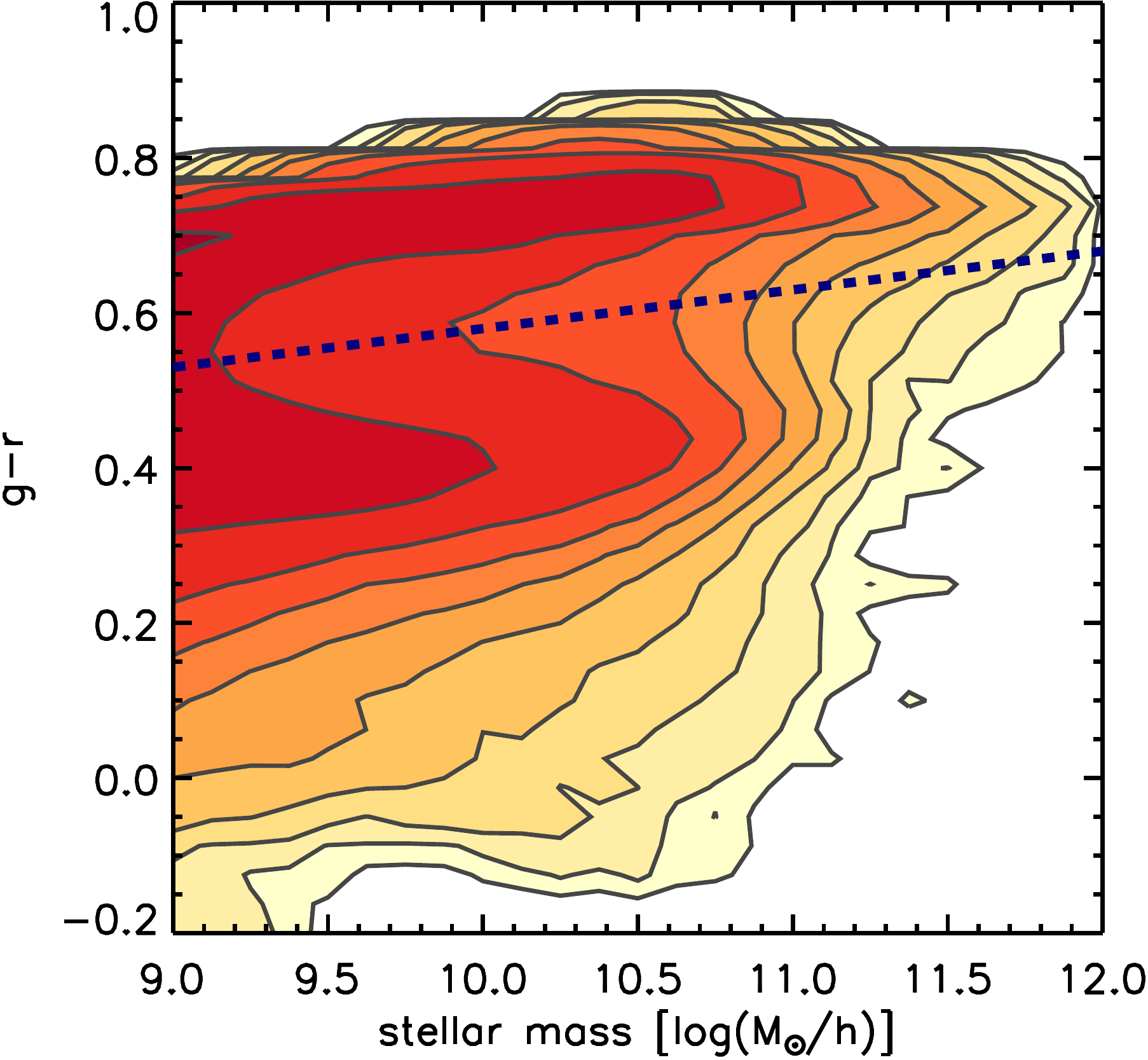}
\caption{The color-mass relation at $z=0$ in the {Guo} {et~al.} (2011) semi-analytical model is shown here as an example. The division between starforming and quiescent galaxies is indicated by the blue dashed line. At higher redshift, this division is adjusted accordingly.}
\label{colormass}
\end{center}
\end{figure}

\begin{figure*}[t]
\begin{center}
\includegraphics[scale=0.4]{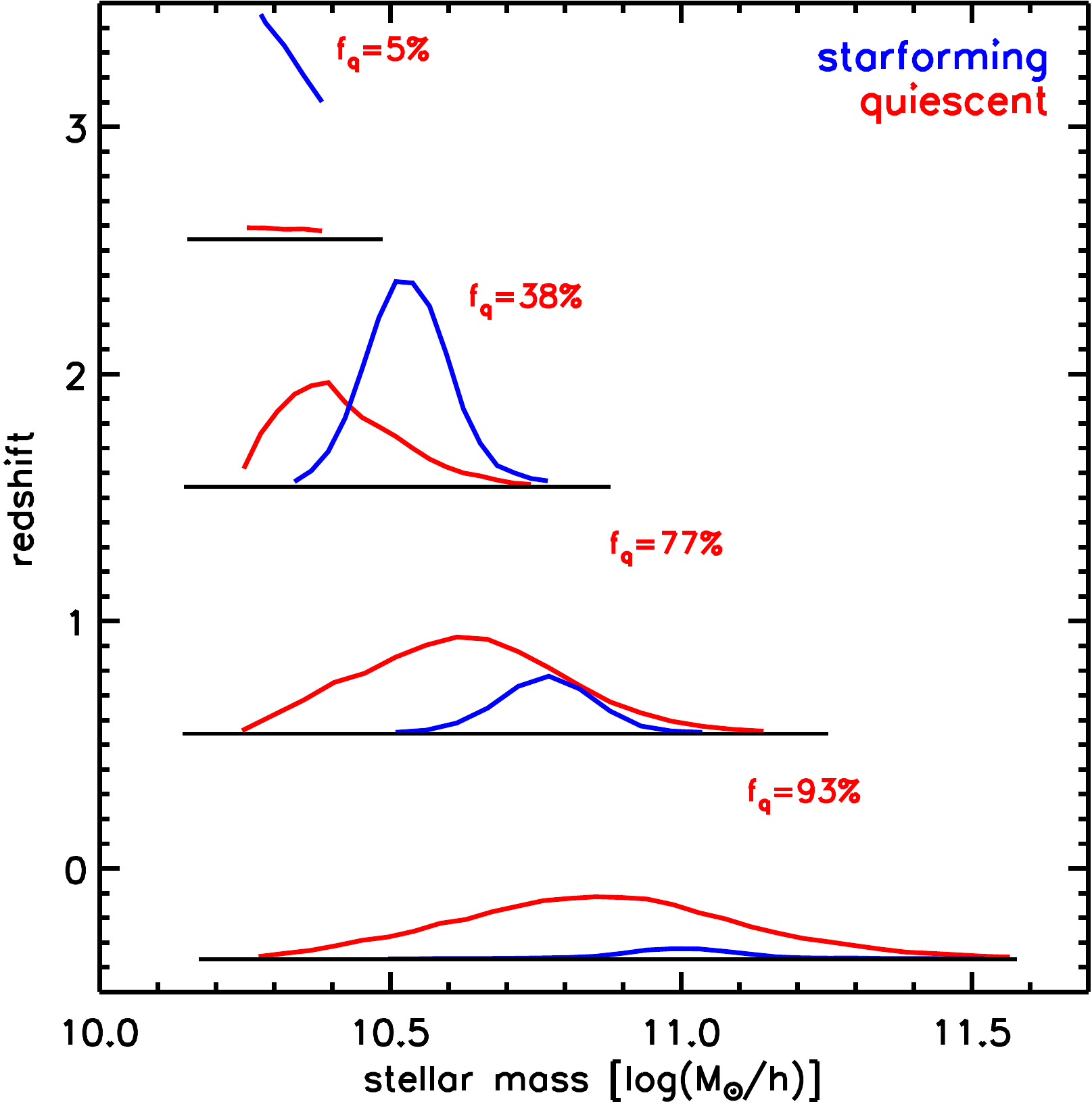}
\includegraphics[scale=0.4]{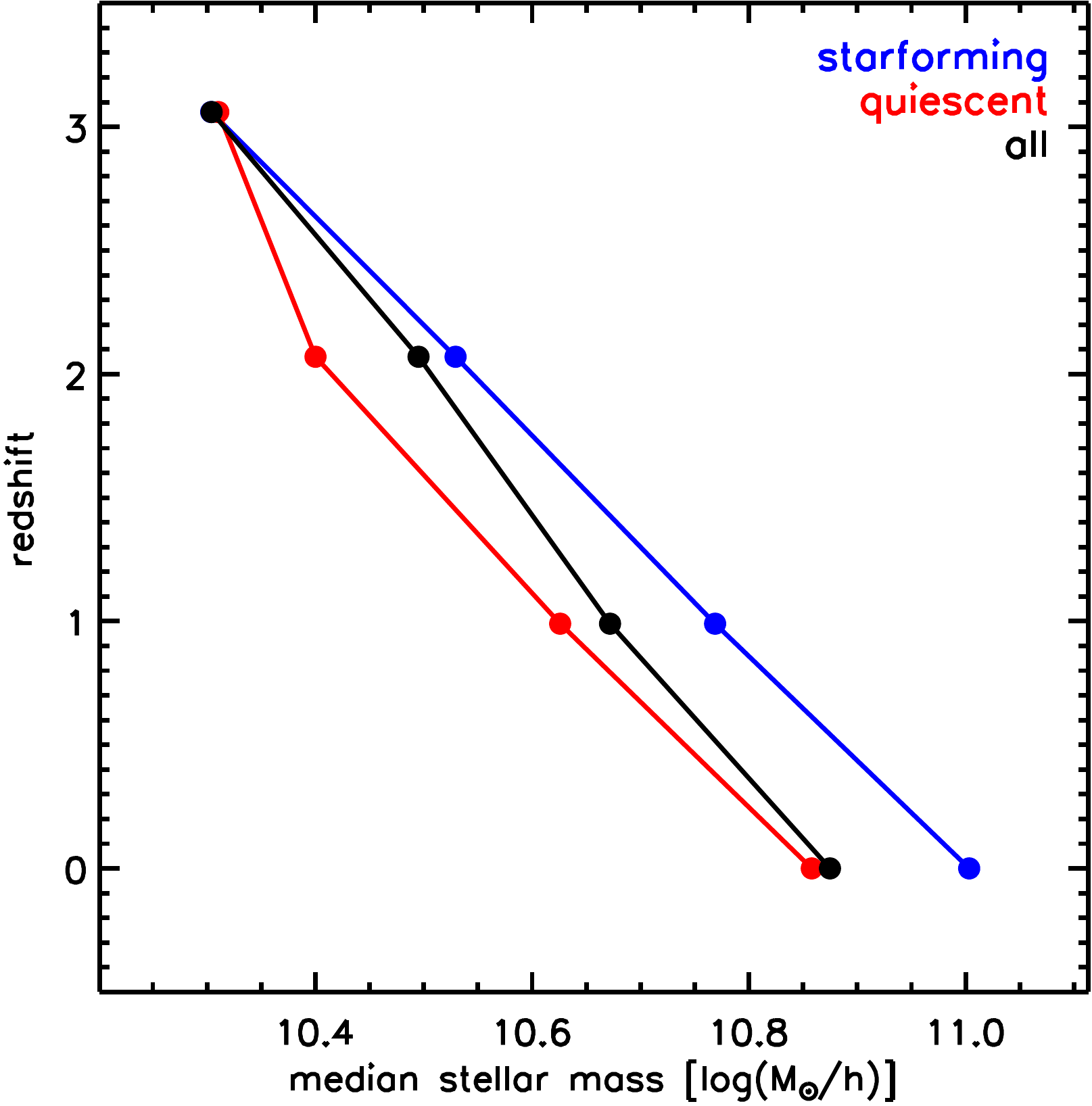}
\caption{The left panel shows the mass distribution of the descendants of the $n=2\times10^{-4}$ Mpc$^{-3}$ number density panel, split into starforming and quiescent galaxies based on a color-mass cut (see Fig. 7). The y-axis of the inset plots is dN/dlog($M_{star}$) with a consistent y-axis from z=2 to z=0: this ensures that the area under the curves accurately tracks the number of galaxies present. The distributions are shown at $z=3.06$, 2.07, 0.99, and 0.00. The right panel shows the evolution of the median mass of the same descendants. The median of the entire sample is shown in addition to as well as the starforming and quiescent subsamples. It is clear that after $z=3$, the median mass of the starforming and quiescent galaxies grow at similar rates.}
\label{massgrowth}
\end{center}
\end{figure*}

In order for number density selection to be effective, the stellar mass growth of galaxies must be a monotonically increasing function of stellar mass. This is true separately for both modes of growth: star formation and merging. However, if star formation is shut down in a certain subset of the population, a discrepancy may arise between the mass growth rate of starforming and quenched galaxies. This will confuse rank order. We investigate the effect by using a color cut to separate quiescent and starforming galaxies at each redshift (see Figure \ref{colormass}) and then plotting the distribution of mass for both populations as a function of redshift for the $n=2\times10^{-4}$ Mpc$^{-3}$ number density slice (see Figure \ref{massgrowth}).

%%%%%%%
%% FIG 8 %%
%%%%%%%

%%%%%%%
%% FIG 9 %%
%%%%%%%

%%%%%%%
%% FIG 10 %%
%%%%%%%

\begin{figure*}[t]
\begin{center}
\includegraphics[scale=0.4]{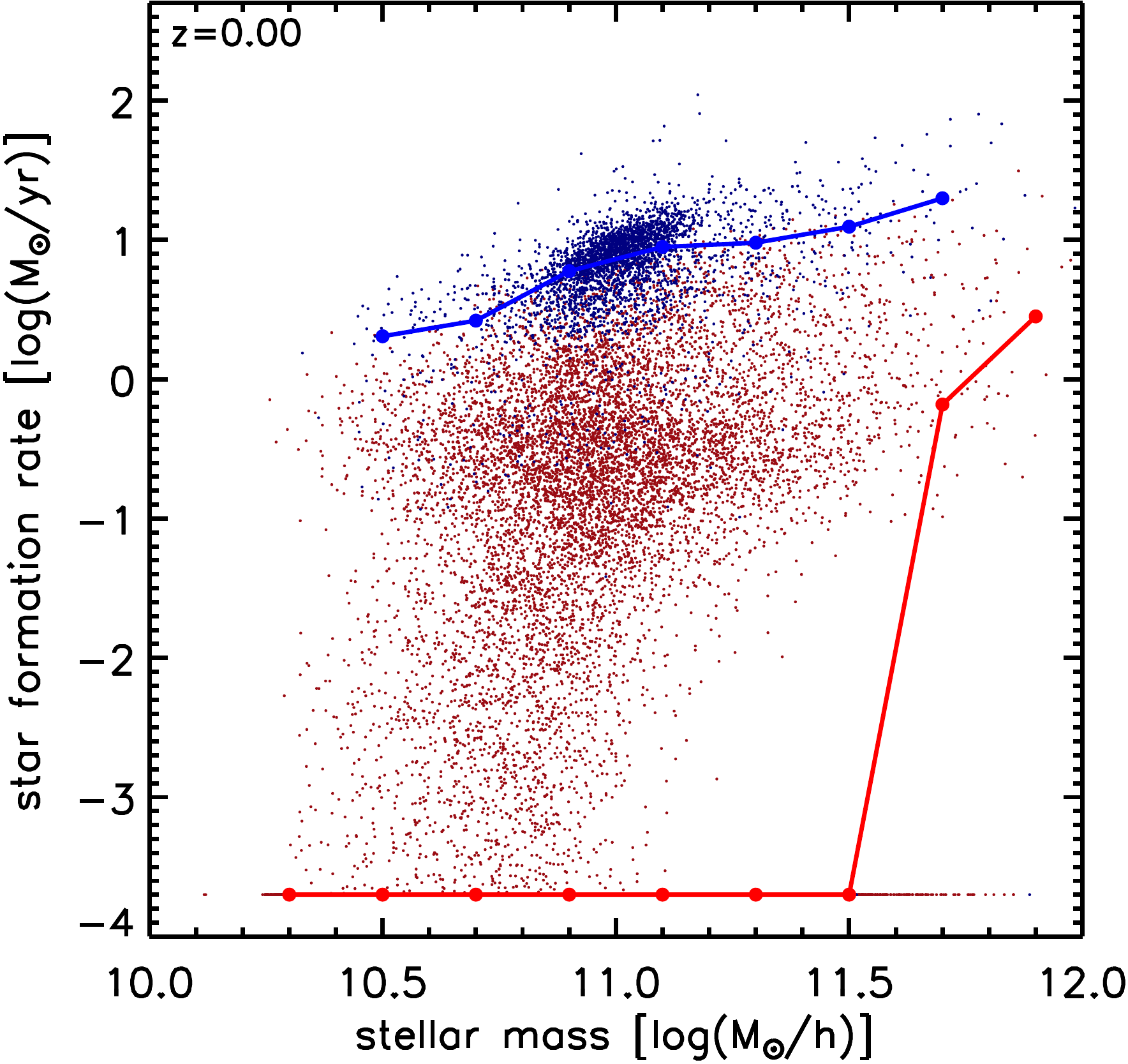}
\includegraphics[scale=0.4]{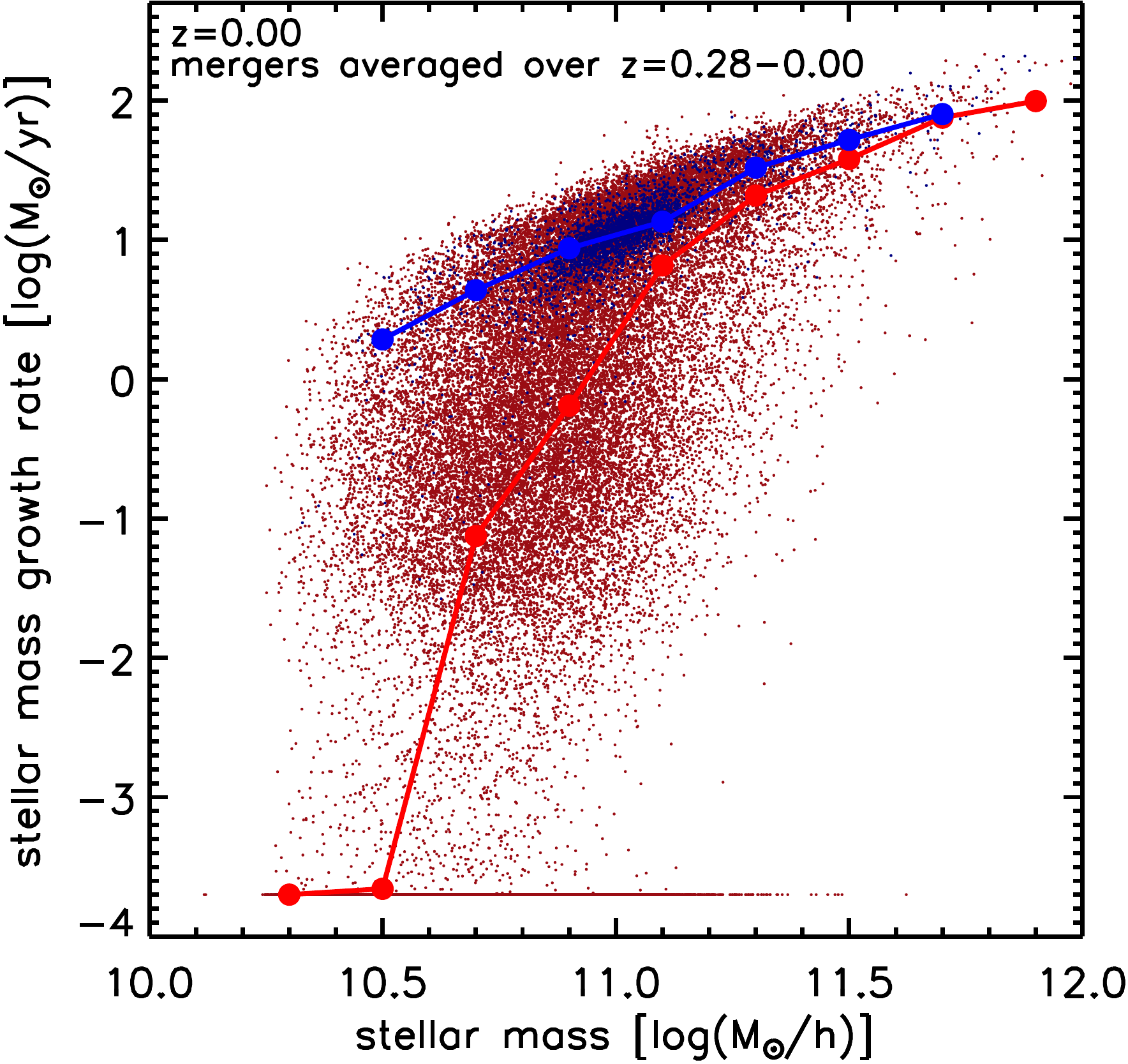}
\caption{Left panel: the descendants from the $n = 2 \times 10^{-4}$ Mpc$^{-3}$ number density selection are shown. Starforming galaxies are blue while quiescent galaxies are red The line is the running median for each sample. The points have a star formation floor at  $\log{(SFR)}$ = -3.7 for illustrative purposes. Right panel: same as the left panel, except now merger growth averaged over time has been added to star formation for both galaxy samples. At the highest masses, the mass growth of quiescent galaxies matches that of star forming galaxies. At lower stellar masses, the mass growth rate diverges, likely because of the predominance of quenched satellites in this mass range.}
\label{sfseq}
\end{center}
\end{figure*}

It is evident from Figure \ref{massgrowth} that there is a clear systematic difference between the mass distribution of starforming and quiescent galaxies at most redshifts. The median mass of starforming descendants  is higher than the median mass of quiescent descendants, except (by construction) the initial selection at $z=3$. After the time between $z=3$ and $z=2$, however, the median mass {\it growth} of quiescent galaxies mirrors that of starforming galaxies. This offset is small (0.15 dex) compared to the spread in stellar mass of the full sample of descendants (full-width half-max $\sim$ 0.5 dex). However, it motivates a closer examination of the processes at work.

A number of low-mass quiescent galaxies are established early on. Due to their low mass, they are likely to be satellites quenched by environmental processes. This low-mass tail is present throughout the simulation, although it is not necessarily composed of the same galaxies in each snapshot. 

The number of quiescent galaxies grows by the quenching of star formation in starforming galaxies. Since starforming galaxies in the sample have a higher median mass than quiescent galaxies, quenching will necessarily grow the upper envelope of the quiescent mass distribution. However, at the highest masses, there exist quiescent galaxies that are more massive than any starforming galaxies. Despite being quiescent, these galaxies must have substantially increased their stellar mass after quenching: the only logical avenue for this is growth via mergers.

This leads to the possibility of three unique galaxy populations, distinguished by their method of stellar mass growth (or lack thereof): starforming galaxies, quenched centrals, and quenched satellites. This is illustrated further in Figure \ref{sfseq}. The left panel plots star formation rate versus stellar mass, with populations split into quiescent (red) and starforming (blue) by a color-mass cut. The running median is shown for both. Clearly, this presents an issue for number density selection, as the quiescent galaxies seem to be adding mass at a much slower rate. However, this only accounts for stellar mass added by star formation.

The right panel of Figure \ref{sfseq} includes stellar mass added by mergers (averaged over $z=0-0.28$) as well as star formation. This is the total mass growth rate as a function of stellar mass. The plot looks markedly different; at masses above $\sim 10^{11}$ M$_{\odot}$, the median mass growth rate for quiescent and starforming galaxies is approximately the same. It is plausible, then, that quenched centrals and starforming galaxies at the same stellar mass are adding stellar mass at the same rate, thus preserving rank order for these galaxies. Quenched satellites remain an issue, as they have ceased adding mass altogether. We caution that these results are sensitive to the details of the quenching model in the SAM and the cosmology of the Millenium Simulation.

%%%%%%%
%% FIG 11 %%
%%%%%%%

\begin{figure*}[h!]
\begin{center}
\includegraphics[scale=0.4]{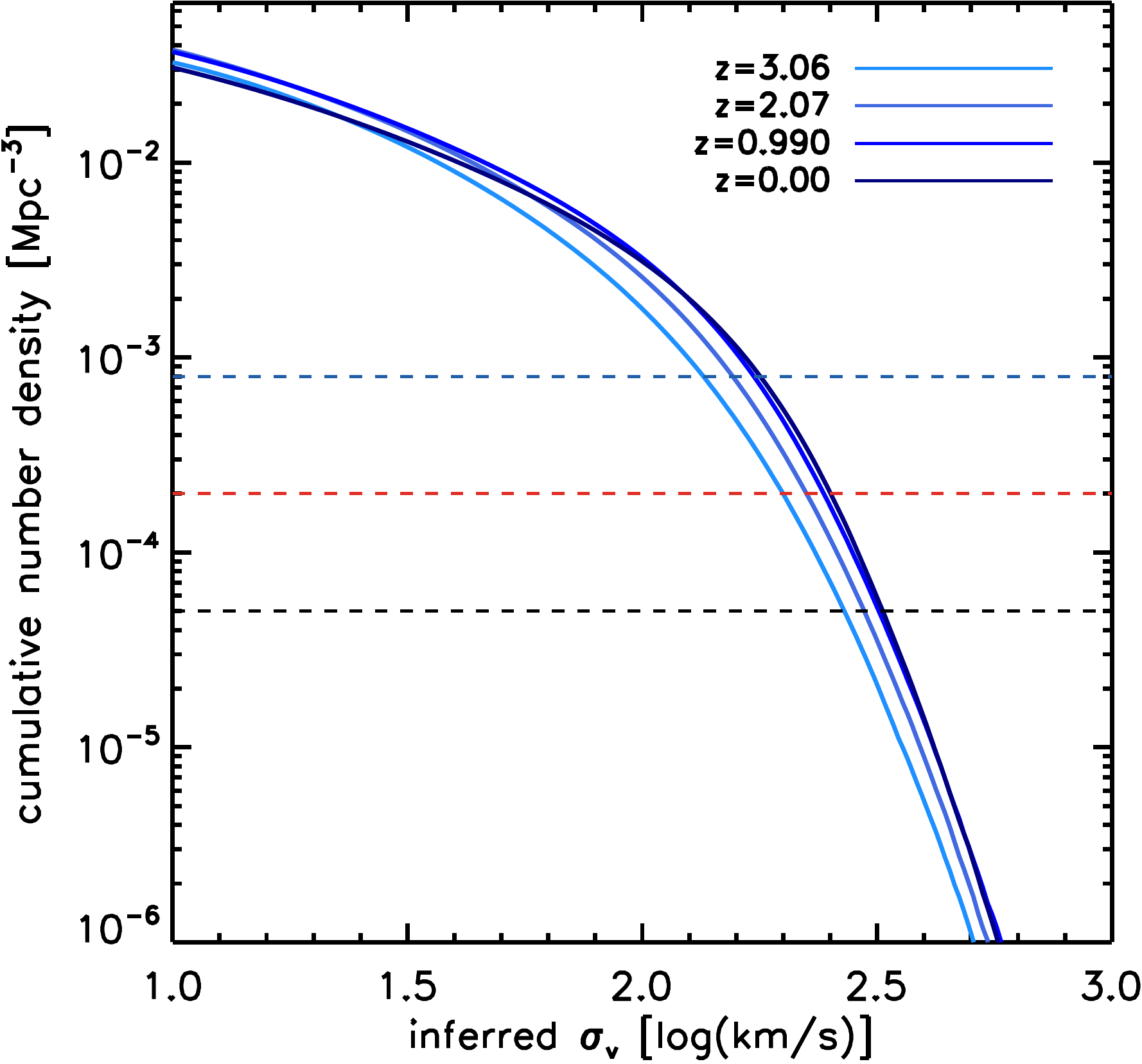}
\includegraphics[scale=0.4]{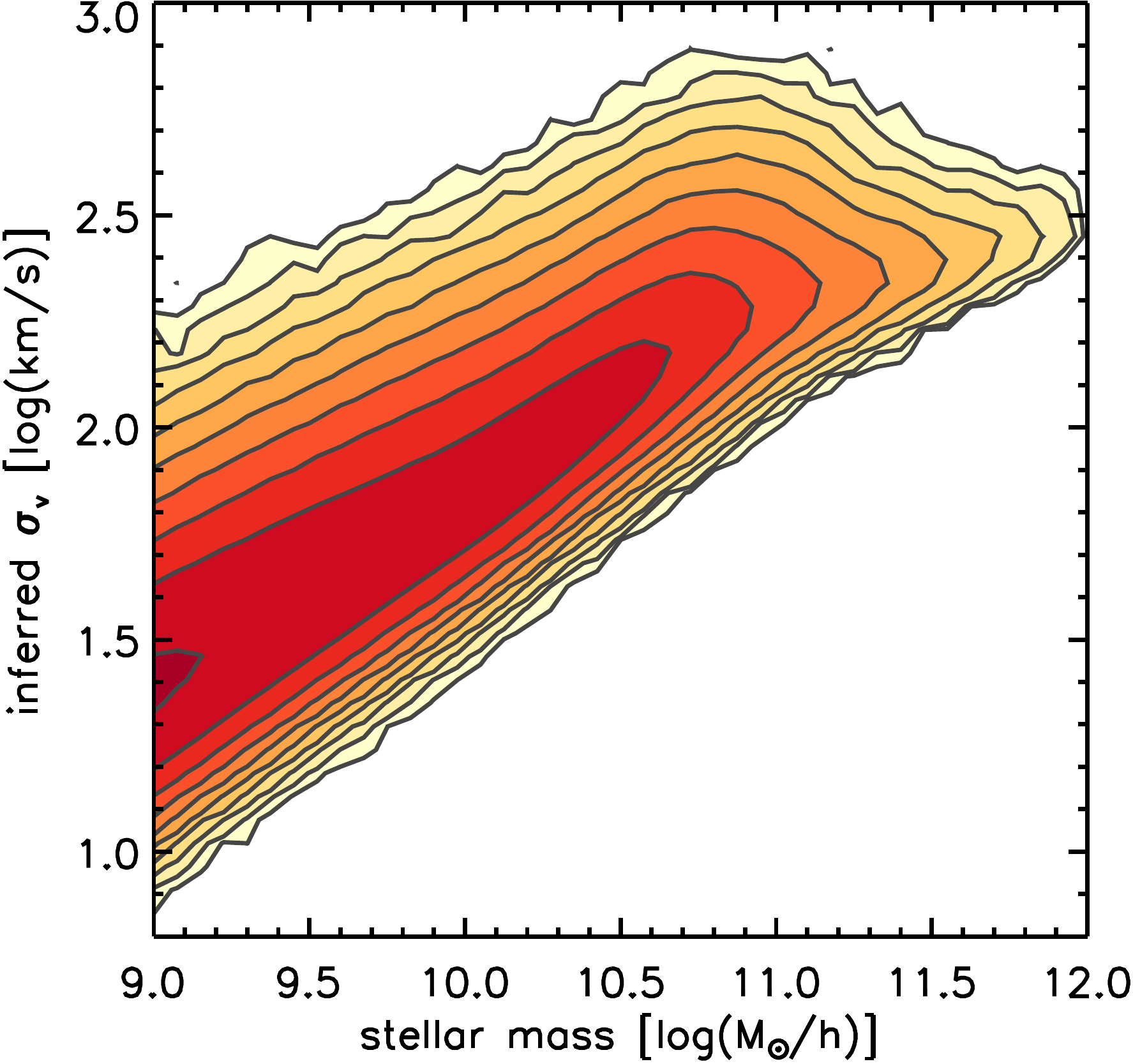}
\includegraphics[scale=0.4]{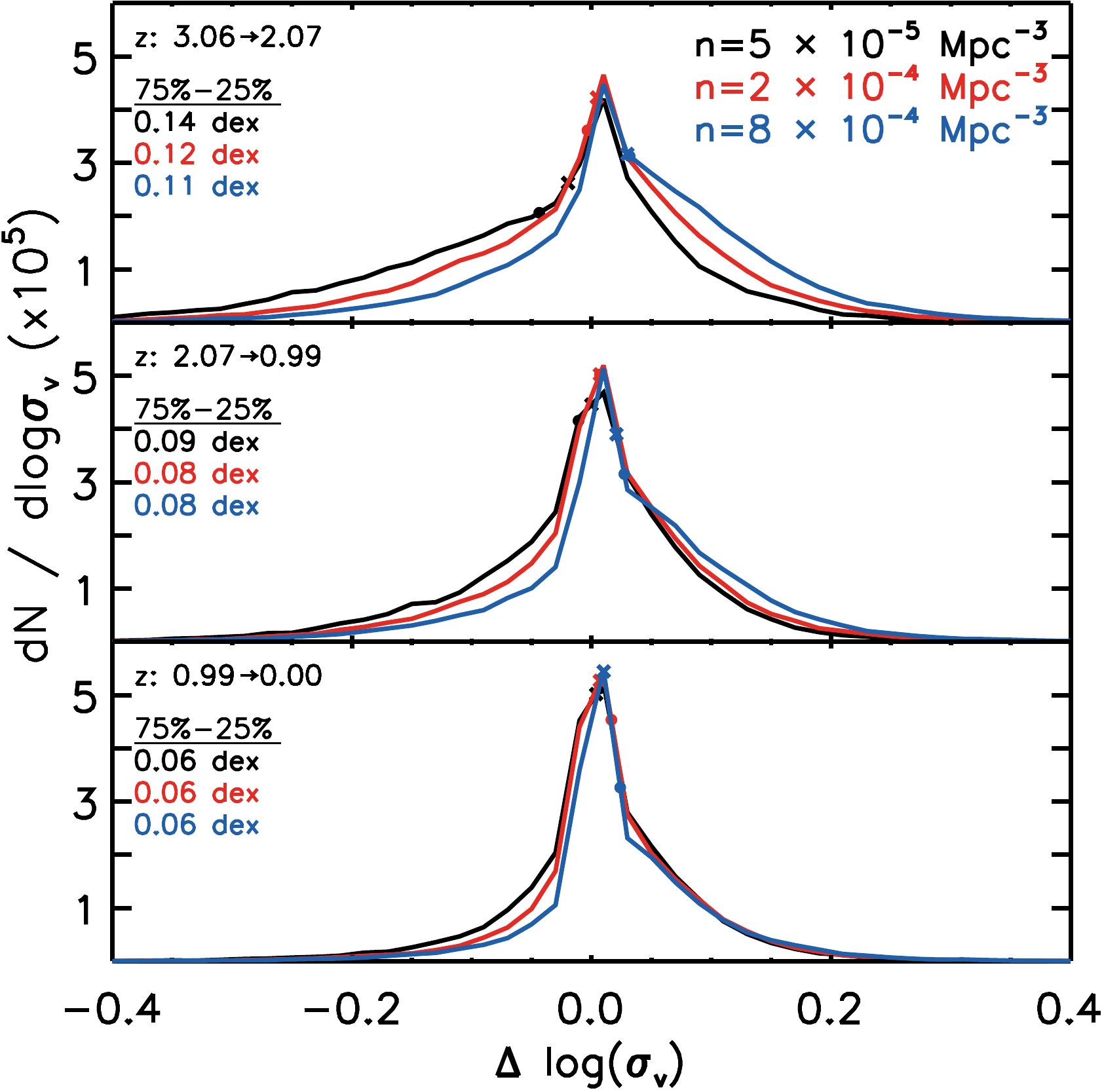}
\includegraphics[scale=0.4]{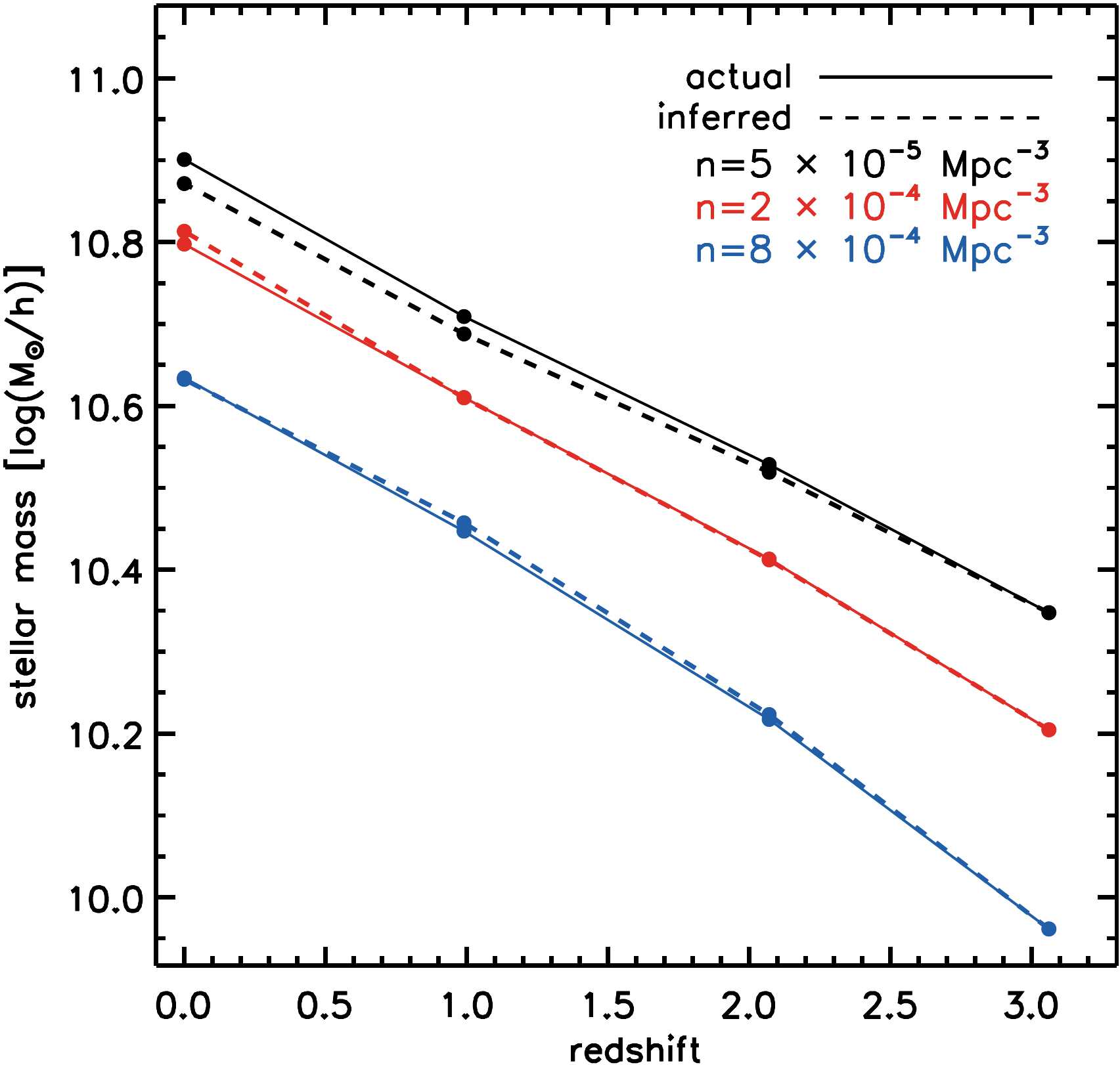}
\caption{Top left panel: the inferred velocity dispersion function for four redshift slices from the Guo et al. (2011) mock galaxy catalogs. The inferred velocity dispersion function is remarkably stable over the whole redshift range. Top right panel: The relationship between inferred velocity dispersion and stellar mass at $z=0$ from the Guo et al. (2011) mock galaxy catalogs. There is no simple one-to-one correlation that would make dispersion selection a trivial version of stellar mass selection. Lower left panel: Tracing the evolution of the descendants of galaxies from the initial selection at z=3.06. The panels show the distribution of evolution in inferred velocity dispersion for individual galaxies. The average is marked with a circle, while the median is marked with a cross. The difference between the 75th quartile and the 25th quartile is shown in the upper left of each panel. The $n=5\times10^{-5}$ Mpc$^{-3}$ descendants are multiplied by a factor of two for comparison, to compensate for the smaller initial bin size. Note the stability of inferred velocity dispersion for individual galaxies as well as the whole distribution. Lower right panel: The median stellar mass growth as a function of number density and redshift, with no adjustments applied. Dispersion selection traces the median stellar mass with remarkable accuracy: compare to Figure \ref{simpmass}.}
\label{dispersion}
\end{center}
\end{figure*}

%%%%%%%%%%%%%%%%%%%%%%%%%%%%%%%%%%%%%%%%
%%%%%%%%%%% INFERRED VELOCITY DISPERSION %%%%%%%%%%
%%%%%%%%%%%%%%%%%%%%%%%%%%%%%%%%%%%%%%%%

\section{Ordering Galaxies by Inferred Velocity Dispersion}\label{sec:disp}
Throughout this paper we have used stellar mass to order galaxies. Another promising approach to number density selection involves ordering galaxies by their stellar velocity dispersion. Velocity dispersion is a measurement of orbital energy and gravitational potential depth. There is evidence from both observations and simulations that velocity dispersion may be a more stable quantity than stellar mass ({Loeb} \& {Peebles} 2003; {Franx} {et~al.} 2008; {van Dokkum} {et~al.} 2010; {Bezanson} {et~al.} 2011; {Oser} {et~al.} 2012). Additionally, velocity dispersion has been proven to be a better predictor than stellar mass of specific star formation rate and color ({Franx} {et~al.} 2008; {Wake}, {van Dokkum}, \&  {Franx} 2012b), galaxy clustering properties ({Wake}, {Franx}, \& {van  Dokkum} (2012a)-- but see also {Li}, {Wang}, \& {Jing} (2012)), and star formation rate ({Bezanson}, {van Dokkum}, \&  {Franx} 2012) over a wide range of redshift. This motivates an attempt to track galaxies by their stellar velocity dispersion, rather than their stellar mass.

Since velocity dispersion is not reported in the Guo et al. (2011) galaxy catalogs, it must be inferred from stellar mass and size. We use the following formula to calculate stellar velocity dispersion:
\begin{equation}
\sigma^2_v = \frac{GM_*}{5R}
\end{equation}
This simple estimate is derived from the scalar virial theorem (see, e.g., {Cappellari} {et~al.} (2006)). 
$M_*$ is the total stellar mass, $G$ is Newton's gravitational constant, and $R$ is the half-mass radius. The {Guo} {et~al.} (2011) galaxy catalogs provide the exponential scale radius of the disk, which is converted into the half-mass radius via $R_{\mathrm{half-mass}} = 1.678R_{\mathrm{exp}}$. The three-dimensional half-mass radius of the bulge is converted into a two-dimensional projected half-mass radius following work in {Ciotti} (1991) by assuming an $n=4$ S\'{e}rsic profile ({Sersic} 1968). These two radii are combined into a projected half-mass radius of the entire galaxy:
\begin{equation}
R_{tot} = \frac{M_{bulge}R_{bulge} + M_{disk}R_{disk}}{M_{bulge}+M_{disk}}
\end{equation}
Assuming perfect photometry and a spatially homogenous stellar population, this would be the observed effective radius.

The upper left panel of Figure \ref{dispersion} shows the dispersion abundance functions as calculated for galaxies from the {Guo} {et~al.} (2011) SAM. They are remarkably similar across all redshifts. While a stable velocity dispersion abundance function implies a stable velocity dispersion for individual galaxies, it is not conclusive. The rate of change of inferred velocity dispersion for individual galaxies is shown in the lower left of Figure \ref{dispersion} as well. Comparison to the same plot for stellar mass (Figure \ref{specevo}) shows that dispersion is considerably more stable: for example, most galaxies change less than 0.06 dex in dispersion from $z=1$ to $z=0$, whereas the spread in stellar mass growth in the same time period is 0.24 dex. The upper right panel of Figure \ref{dispersion} shows the relation between stellar mass and velocity dispersion for all galaxies in the Millenium Simulation, establishing that the described dispersion selection is unique and not a simple function of stellar mass.

Finally, in the lower right panel, the mass evolution of galaxies at a constant cumulative inferred dispersion is compared to the mass evolution of descendants of the original dispersion selection at $z=3$. This is analogous to Figure \ref{simpmass}, except now we are selecting galaxies at each snapshot in redshift by their velocity dispersion and comparing their stellar masses. Constant cumulative dispersion selection predicts the median stellar mass to within 0.05 dex without correction. The total mass evolution of the descendants is $\sim 0.6$ dex, similar to that of the stellar mass selection over the same redshift range. This implies dispersion selection traces the stellar mass evolution to a much higher accuracy. Also notable is the lack of a systematic difference between the inferred and true descendants at the higher number densities. There is a hint of a systematic discrepancy at the lowest number density, but it is small.

Inferred velocity dispersion is determined from the quantity $M_*/R$. Thus, the applicability of the results shown here to the real Universe is dependent on the accuracy of the evolution of $M_*/R$ in the SAM. While the stellar mass evolution has been analyzed extensively in the previous section, it has been shown that the bulge and disk sizes calculated in the {Guo} {et~al.} (2011) galaxy catalogs do not match observations. This will be discussed further in Section \ref{sec:size}. In spite of this, the stability of stellar velocity dispersion with respect to merging, quenching, and satellites/centrals indicates this may be the most promising approach to number density selection.

%%%%%%%%%%%%%%%%%%%%%%%%%%%%
%%%%%%%%%%% DISCUSSION %%%%%%%%%%
%%%%%%%%%%%%%%%%%%%%%%%%%%%%

\section{Discussion: Can Number Density Selection Be Tested in a SAM?}\label{discussion}
We have shown that tracking galaxies at a constant cumulative number density in stellar mass reproduces the descendant mass evolution to within 40\% over a wide range in number densities. Correcting for the scatter in growth rates and merging galaxies reduces this error to 12\%, with the remaining error presumably coming from inaccuracies in the correction method and systematics introduced by satellite quenching. The functionality of number density selection and its dependence on the merger rate, the quenching mechanism(s), and the scatter in stellar mass growth rates are results that are independent of the SAM framework. Since the details of semi-analytical models and N-body simulations do not perfectly mirror the real Universe, however, any quantitative results derived from them must be subject to scrutiny before they are applied to observational studies.
\subsection{The Millenium Cosmology}\label{sec:cosmo}
One of the known issues with the Millenium Simulation is the cosmology: as discussed in Section \ref{sec:mill}, the Millenium $\sigma_8$ is nearly 4$\sigma$ higher than that from the WMAP7 results ({Komatsu} 2010). This results in a higher clustering amplitude for dark matter than is observed in the real Universe. The projected autocorrelation function of stellar mass for the {Guo} {et~al.} (2011) galaxy catalogs overestimates the clustering amplitude by 10-20 percent on large scales and a factor of two on small scales as compared to SDSS results from {Li} \& {White} (2009).

An inflated $\sigma_8$ parameter will affect the stellar mass growth rate in two ways. First, it will increase the merger rate. Second, it will increase the fraction of quenched satellites at a fixed stellar mass. Both effects mean that number density selection will more accurately trace the median mass evolution of galaxies in the real Universe than in the {Guo} {et~al.} (2011) galaxy catalogs. This is explained below for each effect.

Since mergers are necessarily stochastic events, increasing the merger rate will increase both the scatter in stellar mass growth rates and the fraction of galaxies that have "disappeared" in a merger with a larger galaxy. If two of the three sources of error are then lower in the real Universe, this will only strengthen the efficiency of number density selection. Thus, it is expected that the discrepancy between the inferred and actual stellar mass evolution will be less than 0.15 dex when applied to the real Universe. It should be noted that decreasing the merger rate will also decrease the mass growth rate due to mergers: the remarkable congruence in Figure \ref{sfseq} between the mass growth of quiescent and starforming galaxies may not hold if the merger rate is lowered.

Satellite quenching is a problem for number density selection, as it removes galaxies from the starforming sequence in a stochastic way. Quenched satellites will stop growing in stellar mass, whereas a field galaxy of similar stellar mass will continue to grow: this scrambles the rank order of galaxies. Thus, an increase of the fraction of quenched satellites will induce more confusion in the rank ordering of galaxies. If the real Universe has fewer quenched galaxies than the SAM dictates, then it is reasonable to expect that number density selection will trace the median stellar mass with greater accuracy than in the {Guo} {et~al.} (2011) SAM. 

It is also noteworthy that the mass growth rate will affect the redshift evolution of the stellar mass function. Since the Millenium cosmology results in many galaxies quenching too "early", the semi-analytical model is forced to build up the low-mass end of the galaxy mass function at early times. This results in an excess of low-mass galaxies at high redshift compared to observations-- see {Guo} {et~al.} (2011) for in-depth discussion. This discrepancy between the semi-analytical mass function and the observed mass function is a result of a prescription for $\dot{M_*} (M_*)$ that differs from the real Universe. At least some of the issues causing this discrepancy have been addressed above; however, until semi-analytical models can accurately produce the redshift evolution of the mass function, their quantitative predictions for galaxy stellar mass evolution cannot be fully trusted.

\subsection{Mass Growth of Quiescent Galaxies}
In order for the number density selection scheme to be successful, quiescent galaxies must experience significant mass growth through galaxy-galaxy mergers. This is true in the SAM, as evidenced by Figure \ref{sfseq}. Evidence for the importance of mergers in growing stellar mass comes from studies of cosmological and N-body simulations ({Naab}, {Johansson}, \& {Ostriker} 2009; {Oser} {et~al.} 2010; {Feldmann} {et~al.} 2010; {Lackner} {et~al.} 2012). Additionally, the tidal tails and streams seen in deep imaging of local massive quiescent galaxies support the picture of mass growth via mergers ({van Dokkum} 2005; {Tal} {et~al.} 2009; {Janowiecki} {et~al.} 2010). Thus, it is a reasonable expectation that quiescent galaxies still experience a significant amount of stellar mass growth via mergers. Whether it is sufficient to match the mass growth of starforming galaxies-- as is true in the SAM-- requires a careful observational study that is beyond the scope of this paper.

\subsection{Size Growth and Inferred Dispersion}\label{sec:size}
The efficiency of ordering galaxies by their velocity dispersion, as opposed to their stellar mass (see Section 5), is dependent upon conservation of rank order in $M_*/R$. This means that accurately modeling the size evolution of galaxies within the SAM is important. In the semi-analytical model, disk sizes are set by the angular momentum of their stellar component, which is in turn set by the angular momentum of the gaseous component at the moment it is transformed into stars. The specific angular momentum of gas accreted from the IGM is set to be equal to the specific angular momentum of the halo; gas acquired during mergers has its own angular momentum which is added to the galactic gas. Bulges grow during major mergers, minor mergers, and due to disk instabilities. Their size growth is determined by energy conservation and the virial theorem.

{Guo} {et~al.} (2011) notes there are systematic differences between the resulting distribution of bulge and disk sizes as compared to observations. Both the median bulge and the median disk sizes are too high at low stellar masses, and too low at high stellar masses. Additionally, the simulation scatter is larger than the observed scatter. In particular, the small observed scatter in stellar bulge size with stellar mass is difficult to understand if bulges are built primarily via mergers ({Nair}, {van den Bergh}, \& {Abraham} 2010), as is true for high and low mass galaxies (although not intermediate, Milky Way mass galaxies) in the {Guo} {et~al.} (2011) model. This casts doubt on the accuracy of stellar mass predictions derived from ordering galaxies by their velocity dispersion. 

However, the current model for size evolution incorporates most known important physical processes and should not be dismissed out-of-hand. For example, the velocity dispersion function is believed to be remarkably stable ({Bezanson} {et~al.} 2011), and this stability is reproduced in the model. Additionally, as reflected in the merger model in {Guo} {et~al.} (2011), some physical processes are expected to couple mass and size growth. In this scenario, velocity dispersion may remain a relatively stable quantity. Exploring whether constant number density in velocity dispersion, in stellar mass, or simple constant velocity dispersion is the best approach to tracking galaxies will require a more sophisticated and robust semi-analytical recipe for velocity dispersion, possibly attainable by comparison to hydrodynamical simulations.

\subsection{What Results are Independent of SAM Recipes?}
There are certain features that must exist in a semi-analytical model in order for it to be a reasonable testbed for number density selection. The star formation rate must growth with the stellar mass of a galaxy, and the merger rate must increase monotonically as a function of stellar mass. A quenching model that turns off both massive centrals and satellite galaxies must exist, in parallel with the real Universe. Spontaneous galaxy creation must be a negligible effect at higher stellar mass. These features are all evident in the {Guo} {et~al.} (2011) SAM. This allows us to extract conclusions from the behavior of galaxies in the SAM which may then be applied to observational studies, independent of details in the semi-analytical and numerical modeling. 

The simple techniques used to adjust the inferred mass growth rely only on a few observables: the shape of the stellar mass function, the merger rate as a function of stellar mass, and quantification of the starforming sequence. All of these may be estimated from observations, although someÐ for example, the merger rateÐ are poorly constrained, particularly at high redshift (see, e.g., {Newman} {et~al.} (2012)). Thus, even if the specifics of the semi-analytical model prove to be incorrect, number density selection will still operate within the same framework as described here, making the method can easily generalizable to the real Universe.

The numerical results presented here remain sensitive to the details of the SAM recipes. The slope, scatter, and exact details of the merger rate and the starforming sequence affect the mass evolution as a function of number density. The specifics of the quenching recipe may also prove critical to the efficacy of tracking galaxies at a constant number density. With reasonable prescriptions for these effects, though, the resulting inferred and descendant mass evolution should be good estimates of the truth.

\subsection{Comparison to Tracking Galaxies with a Stellar Mass-Halo Mass Relationship}
The evolution of halo mass is known to high accuracy from numerical simulations. By adopting a redshift-dependent stellar mass-halo mass relationship, one can use the evolution of the halo mass function to infer the evolution of stellar mass (e.g. ({Conroy} \& {Wechsler} 2009; {Behroozi}, {Wechsler}, \&  {Conroy} 2012)). The strength of this approach relative to number density selection is that the growth and merger rates of dark matter halos are known to greater accuracy than the growth and merger rates of galaxies. However, an additional source of scatter is introduced when the stellar mass-halo mass relationship is applied. This scatter is on the order of 0.16-0.2 dex at $z=0$, with the scatter presumably increasing with redshift ({More} {et~al.} 2009; {Reddick} {et~al.} 2012). A comparable analysis to that presented in this paper would be needed is necessary to determine the effects of the scatter in the stellar mass-halo mass relationship on the inferred stellar mass evolution; unfortunately, studies parametrizing $M_*(M_h,z)$ separately for individual halos have not yet been done ({Behroozi} {et~al.} 2012). Pending further study, then, tracking galaxies with information from dark matter simulations may prove to be a viable alternative to number density selection in the future.

%%%%%%%%%%%%%%%%%%%%%%%%%%%%
%%%%%%%%%%% CONCLUSION %%%%%%%%%%
%%%%%%%%%%%%%%%%%%%%%%%%%%%%

\section{Conclusions}
We have investigated the efficacy of tracking galaxies at a constant cumulative number density. We discuss the assumptions inherent in the use of number density selection and which physical processes might violate them. We then use the {Guo} {et~al.} (2011) semi-analytical model based on the Millenium Simulation to examine number density selection in a realistic simulation setting. The technique is demonstrated to be a robust method to track the evolution of an individual galaxy population. Applying the technique to galaxies evolving from $z=3$ to $z=0$, or eleven billion years in time, reproduces the factor of four evolution in stellar mass to within 40\% within the model. Effects that confuse stellar mass rank order have been identified: scatter in stellar mass growth rates, merging galaxies, and quenching. Corrections developed for the first two effects brings the discrepancies down to approximately 12\%. However, a difference in the mass growth rate of quenched galaxies relative to starforming galaxies may produce systematic errors in the assumption of constant rank order. This is sensitive to both the recipes for quenching and the merger rate and thus merits investigation in the real Universe rather than within a semi-analytical model. We also examine ordering galaxies by their inferred velocity dispersion as opposed to their stellar mass. This method reproduces the mass evolution of galaxies from $z=3$ to $z=0$ to within 12\% with no corrections whatsoever, although it is sensitive to the rules governing size evolution in the semi-analytical model.

The ultimate goal of this study is to apply the number density technique to observational studies of the real Universe. However, since it is impossible to observe a galaxy evolve through time, we must calibrate this technique in semi-analytical models. The most obvious concern in this study is that semi-analytical models may not accurately reflect the Universe. Some of the most important discrepancies include the redshift evolution of the mass function and enhanced clustering strength compared to the real Universe. We address these concerns in a number of ways. First, we emphasize the SAM is used primarily as an investigative tool to see how number density selection works and what will disrupt it: the major mechanisms (mergers, the star forming sequence, quenching) affecting this are reproduced in the SAM, regardless of the details of their implementation. Second, the methods developed herein to correct for scatter in growth rates and mergers are based on simple observable quantities like the stellar mass function and the merger rate, and may safely be generalized to the real Universe. Finally, it is shown that the known discrepancies between the semi-analytical model and the Universe act to {\it decrease} the errors in the observational inferred mass evolution relative to those in the SAM. Thus, the errors quoted here should be taken as upper limits to errors that would arise in an analogous study of the real Universe.

Linking descendant and progenitor galaxies at a constant number density is thus shown to be a considerable improvement over previous techniques, and opens the exciting potential for using observations to track changes in individual galaxy populations over long periods of cosmic time.

\acknowledgements

We thank David Wake for his help in the early stages of this project. Support from grant HST-GO 12177.01 is gratefully acknowledged. The Millennium Simulation databases used in this paper and the web application providing online access to them were constructed as part of the activities of the German Astrophysical Virtual Observatory.

\end{document}